\pdfoutput=1

\documentclass[useAMS,usenatbib]{mn2e}

\usepackage{graphicx}
\usepackage{dcolumn}
\usepackage{bm}
\usepackage{amsmath}
\usepackage{amsfonts}
\usepackage{psfrag}
\usepackage{here}
\usepackage{rotating}
\usepackage{fixltx2e}
\usepackage{float}

\title[The Fundamental Plane of Black Hole Activity in the Optical Band]{The Fundamental Plane of Black Hole Activity in the Optical Band}
\author[P. Saikia et al. 2014]{Payaswini Saikia$^{1}$\thanks{E-mail:
p.saikia@astro.ru.nl}, Elmar K\"{o}rding$^{1}$ and Heino Falcke$^{1,2,3}$\\
$^{1}$Department of Astrophysics/IMAPP, Radboud University, Nijmegen, P.O. Box 9010, 6500 GL Nijmegen, The Netherlands\\
$^{2}$ASTRON, Oude Hoogeveensedijk 4, 7991 PD Dwingeloo, The Netherlands\\
$^{3}$Max-Planck-Institut f\"{u}r Radioastronomie, auf dem H\"{u}gel 69, 53121 Bonn, Germany}

\begin{document}

\date{Received: 2015 March 25. Accepted: 2015 March 30.}

\pagerange{\pageref{firstpage}--\pageref{lastpage}} \pubyear{2015}

\maketitle

\label{firstpage}


\begin{abstract}
Black hole accretion and jet formation have long been thought to be scale invariant. One empirical relation suggesting scale invariance is the Fundamental Plane of Black Hole activity, which is a plane in the space given by black hole mass and the radio/X-ray luminosities. We search for an alternative version of this plane using the luminosity of [OIII] emission line instead of X-ray luminosity. We use a complete sample of 39 supermassive black holes selected from the Palomar Spectroscopic Survey with available radio and optical measurements and information on black hole mass. A sample of stellar mass X-ray binaries has also been included to examine if physical processes behind accretion is universal across the entire range of black hole mass. We present the results of multivariate regression analysis performed on the AGN sample and show that the sample stretches out as a plane in the 3D logarithmic space created by bolometric luminosity, radio luminosity and black hole mass. We reproduce the established Fundamental Plane of Black Hole activity in X-rays. We show that this plane can be obtained with the supermassive black hole sample alone and the X-ray binaries agrees to the found relation. We also discuss radio loudness of various classes of low-luminosity AGN in view of our fundamental plane.
\end{abstract}



\section{Introduction}

Accretion physics is thought to scale globally across black holes of different mass scale $-$ from millions of solar masses Supermassive black holes (SMBH) to $\sim$10 solar masses stellar X-ray binaries (XRBs).

Many theoretical arguments and observational evidences have been put forward to support scale-invariance of black hole accretion and relativistic jet physics. The amount of synchrotron radiation emitted from a scale invariant jet is theoretically shown to depend both on the black hole mass and the accretion rate \citep[eg.][]{fb95,hs03}. The observational evidence supporting the same is the fundamental plane of black hole activity connecting XRBs and AGN. It is a non-linear empirical correlation in the space given by the black hole mass, the radio luminosity and the X-ray luminosity with high statistical significance \citep{m03, f04}. The radio luminosity is used in the fundamental plane relation as a probe for the AGN jet \citep{bk79} while the X-ray emission is taken to be a tracer for accretion rate.

In this study, we use the forbidden [OIII] emission line luminosity as an indirect tracer of accretion rate instead of X-ray luminosity and re-establish the fundamental plane relation for black hole activity. Generally, X-ray luminosity is a better proxy for the bolometric luminosity and accretion rate, but they are costly to obtain and hence are not easily available. The [OIII] luminosity, on the other hand, can be measured by ground-based observations and hence is easily available. In addition, as the Narrow Line Region extends far beyond the central source, torus obscuration of the [OIII] emission line is minimized compared to X-ray emission. Moreover, spectral index variations are less of an issue while using the [OIII] line luminosity. In general, [OIII] lines are strong, easy to detect and are known to be relatively weak in metal-rich, star-forming galaxies. Hence, the emission line [OIII]$\lambda$5007 (2s\textsuperscript{2}2p\textsuperscript{2} \textsuperscript{1}D\textsubscript{2} - 2s\textsuperscript{2}2p\textsuperscript{2} \textsuperscript{3}P\textsubscript{2}) is commonly used as a surrogate for the bolometric luminosity \citep[eg.][]{h04,fmb95} and as a tracer of the nuclear luminosity \citep[eg.][]{k03}. Using this [OIII] emission line luminosity as a proxy for the accretion rate and the 15 GHz radio luminosity as the tracer for jet emission, we have examined the disc-jet connection in black holes of different masses by investigating their correlations. Later we convert the [OIII] luminosities of the supermassive black holes to X-ray and Bolometric luminosities in order to compare them with the smaller mass X-ray binaries. 

We use a complete sample of 39 supermassive black holes and a selected sample of the best-studied stellar mass X-ray binaries for this work. The sample is described in Section 2. In Section 3, we present multivariate regression analysis results of the supermassive black hole plane in [OIII]. We reproduce the fundamental plane of black hole activity in Section 4 and introduce the bolometric fundamental plane in Section 5. Finally in Section 6, we discuss our results and present the conclusions of this study in Section 7.


\section[]{Sample Selection}

\subsection{Supermassive black holes}

To study the general properties of accretion, it is preferred to have a complete sample of galaxies covering a broad range of spectral types, luminosities and morphological types. For the entire sample, information on velocity dispersion, radio luminosity and [OIII] line luminosity are required to estimate the black hole mass, jet emission and accretion rate respectively. The best currently available complete sample in the northern hemisphere is the Palomar Spectroscopic Survey \citep{h95}, comprising the nuclear region of 486 nearby galaxies with B$_\tau <$ 12.5 mag. The SMBH sample used for this study has been extracted from this Palomar Survey and hence is mainly targeted on the low-luminosity AGN.

Several multi-wavelength surveys have been performed on the Palomar Sample. Radio luminosities for this study, which are needed as an estimate for the jet emission, have been taken from the high resolution radio survey of all LLAGNs and AGNs in the Palomar sample, presented by \cite{n05}. From the Palomar survey, \cite{n05} selects all 403 galaxies with nuclear emission lines - 206 of these nuclei have H II region type spectra, and the rest 197 are the AGN.They report 197 VLA 15 GHz and 44 VLBI 5 GHz flux measurements. For this analysis, we need the nuclear radio luminosity and hence we need the flat-spectrum radio cores in our galaxies. Therefore, we use only the VLA data as the VLA survey gives higher frequency luminosities and hence is designed to pick up galaxies with a flat spectrum radio core.

The [OIII] line luminosities, used in the study as a tracer of the accretion rate, have been taken from \cite{h97}, where spectroscopic properties and parameters for 418 AGN and their host galaxies are presented. It is important to note that the [OIII] emission line can be attenuated by dust within the host galaxy. The observed [OIII] luminosity can be corrected for this extinction by using the Balmer decrement  
\begin{equation}
L^{c}_{OIII} = L_{OIII} \left(\frac{(H\alpha/H\beta)_{obs}}{3.0}\right)^{2.94} ,
\end{equation}
where the intrinsic Balmer decrement is taken to be 3.0 \citep[eg.][]{of06}.
The black hole mass estimate can be obtained through the established empirical relation between black hole mass and central stellar velocity dispersions. For this study, we use the stellar velocity dispersions presented in \cite{h09} to derive the black hole masses. These velocity dispersions are calculated using the same spectra. The equation we use to determine mass from velocity dispersion is \citep{m11}
\begin{equation}
log \frac{M_{BH}}{M_\odot} = 8.29 + 5.12 log_{10}  \frac{\sigma}{200km/s} .
\end{equation}

We restrict our final sample to the galaxies which have available data in 15 GHz radio luminosity, [OIII] emission line luminosity and black hole mass; yielding a sample of 101 galaxies. This restriction introduces observational bias in the sample. In order to properly constrain the parameters of the proposed correlation, we exclude the upper limits from our final dataset. We later perform partial Kendall $\tau$ correlation test to check for the significance of the plane and find that the underlying correlation exists even in the presence of the upper limits.

Removing the upper limits reduces the sample size to 39 supermassive black holes, comprising 20 LINERS, 12 Seyferts and 7 Transition galaxies. The final sample can be morphologically classified as 7 Ellipticals, 18 Spirals, 1 Irregular and 13 Lenticulars. This is the complete sample of AGN from the optically-selected Palomar survey of all northern galaxies, showing AGN-like spectra and having detected radio and [OIII] line luminosities.

\subsection{Stellar black holes}

A homogeneous sample of black holes covering the entire black hole mass range is required to study general properties of accretion. Hence, we include stellar mass galactic black holes to our sample of supermassive black holes. Stellar mass XRBs are classified into different states according to their accretion - the low hard, the high soft and the intermediate states \citep{rm06}. The radio spectrum in the low/hard state is consistent with the spectra of a steady jet \citep{fe04} while radio emission seems to be quenched in the high/soft state \citep{f99,c00}.

X-ray Binaries in the low/hard state follow a universal correlation between the radio and X-ray luminosity of the form $L_R$ $\propto$ $L^{0.7}_X$ \citep{gfp03}, dominated by the observations of GX 339-4 and V404 Cyg. For our study, we take a selected sample of the best-studied X-ray Binaries in the hard state. This study does not use the 'radio-quiet' XRBs forming the Ôoutlier trackÕ of the universal radio/X-ray correlation \citep[eg.][etc.]{c11,g12}.\\

\noindent \textbf{GX 339-4:} Data presented in \cite{c03} as 88 quasi-simultaneous radio and X-ray observation from a long-term campaign of GX 339-4 in the low/hard state, has been used for this study. GX 339-4 is believed to have a black hole with a mass $>$ 5.8 $M_\odot$ \citep{hy03}. The distance to GX 339-4 is taken to be 8 Kpc \citep{z04}. 

\noindent \textbf{V404 Cyg:} We use the VLA radio and \textit{Chandra} X-ray observations of V404 Cyg in the hard state, as reported in \cite{c08} for this study. The distance to this source is taken to be 3.5 Kpc \citep{z99} while the black hole mass is taken as 10 $M_\odot$ \citep{s96}.

\noindent \textbf{XTE J1118+480:} The X-Ray transient XTE J1118+480 at it's hard state is also included in the XRB sample. The radio and X-ray luminosities are directly taken from the compilation in \cite{m03}.

\noindent \textbf{A0620-00:} Finally, we also include the simultaneous \textit{Chandra} X-ray and VLA radio observation of A06200-00 in the hard state, as reported in \cite{g06}. This source has a black hole of mass 11.0$\pm$1.9 $M_\odot$ \citep{g01} and lies at a distance of 1.2$\pm$0.4 Kpc \citep{sn94,jn04}.


\section{Fundamental plane in [OIII]}

The final sample of supermassive black holes stretches out in a plane in the three-dimensional logarithmic space defined by the 15 GHz radio luminosity, [OIII] line luminosity and black hole mass. The resulting plane is obtained with some scatter owing to various measurement errors and intrinsic variability of sources.

\subsection{Multivariate Regression Analysis}

We perform a Multivariate Correlation Linear regression analysis on the data to estimate the parameters for the plane. Standard chi-square fits used for analysis yield asymmetric results as it can not consider the scatter in all variables \citep[eg.][]{fv98}. Hence, we use the modified chi-square estimator known as merit function, as uncertainties are present in all the required measurements. The merit function is defined as
\begin{equation}
\chi^2(a,b) = \sum\limits_i \frac{(y_i - b -  \sum\limits_j a_jx_{ij})^2}{\sigma_{yi}^2 + \sum\limits_j (a_j \sigma_{x_{ij}})^2} ,
\end{equation}
where $\sigma _{x_{ij}}$ and $\sigma _{yi}$ are the respective uncertainties. Here \textit{y$_i$} denotes the [OIII] line luminosities, \textit{x$_{1j}$} the radio luminosities at 15 GHz and \textit{x$_{2j}$} the black hole masses. The linear regression coefficients a$_j$ and the zero intercept b are the unknown parameters that can be found by minimizing $\chi^2$.

It is not possible to minimize the merit function analytically as the equation is nonlinear in a$_j$. But we can analytically solve for the constant b and then a simple numerical optimization routine can be used to find the parameter a$_j$s.

\subsection{Error budget}

To correctly extract the parameters of the plane, it is crucial to estimate the errors cautiously for each variable. It has been observed that the resulting parameters of the fundamental plane correlation depend strongly on the choices of assumed uncertainties in the data \citep{k06}.

Uncertainty in luminosity depends on both flux and distance measurements. Errors in radio and optical fluxes are different for each data point, and are usually very low - less than 10\%. For this analysis, we use a typical value of 0.05 dex as the uncertainty in flux. Errors in distance measurement typically range from 0.1 to 0.4 dex. We adopt an uniform value of 0.15 dex for our distance uncertainty.

In addition to these, we include an intrinsic scatter term whose exact magnitude is chosen to ensure that the reduced merit function is unity. For the SMBH plane, we get an intrinsic scatter of 0.35dex. It gets further reduced to 0.2 dex for the combined plane obtained with the complete black hole sample of SMBH and XRBs. This scatter term takes into account various factors like source peculiarities, non-simultaneous measurements of radio and optical flux of AGN, effects of spin, beaming statistics, absorption etc.

We estimate the SMBH masses from the available velocity dispersions using the M-$\sigma$ relation. \cite{mf01} give an absolute scatter of 0.34 dex for this relation, which can be used as a measure of the uncertainty for the SMBH mass estimate. 

\subsection{Correlation Tests}


\begin{figure}
\vspace{3.02pt}
 \includegraphics[width=86mm]{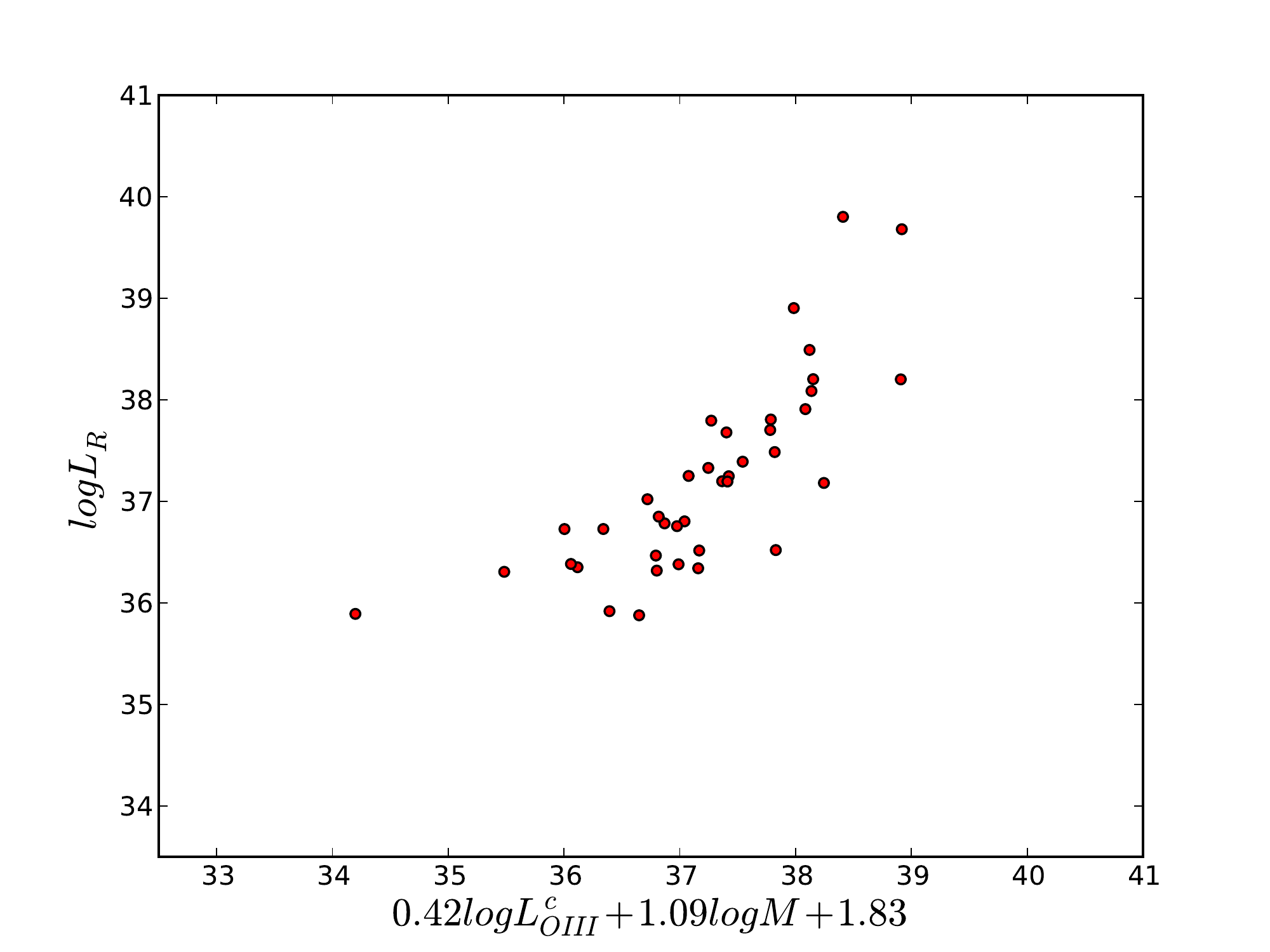}
 \caption{Projection of the Optical Fundamental Plane with Supermassive Black Holes. Luminosities are given in erg/s while the masses are in the unit of solar mass.}
\end{figure}

We perform the Kendall Tau Correlation Test on our SMBH sample to statistically verify the significance of the plane. The Kendall tau value is a measure of the correlation between the specified measurements. For the L$_R$-L$_{OIII}$-M plane defined by the sample, we get a Kendall Tau coefficient value of $\tau = 0.67$ (with the probability for null hypothesis as P$_{null} \simeq 1.3 \times 10^{-9})$. For the plane obtained with extinction-corrected [OIII] luminosities (see Fig 1), we obtain $\tau = 0.66$ (with P$_{null} \simeq 2.8 \times 10^{-9})$.

It is also important to statistically check for spurious effects in Luminosity-Luminosity plots due to their common dependence on distance. Kendall Tau Partial Correlation analysis was performed with distance as the third variable. The correlation was found to be real even after taking into account the large range of distance ($\tau = 0.48$, with P$_{null} \simeq 3.9 \times 10^{-3})$.

As mentioned, the sample size substantially increases from 39 to 101 galaxies, if upper limits are included in the dataset. In order to assess the significance of the apparent correlations in the presence of upper limits, we have adopted the method proposed by \cite{as96}, using which a Partial Kendall Tau correlation test can be performed in the presence of censored data. Applying this test to our data, we find a Partial Kendall $\tau$ value of 0.36 (with P$_{null} \simeq 4.4 \times 10^{-12})$ showing that although the significance of the plane reduces while taking into account the upper limit data, a real underlying correlation indeed exists.


\begin{figure}
 \includegraphics[width=87.2mm]{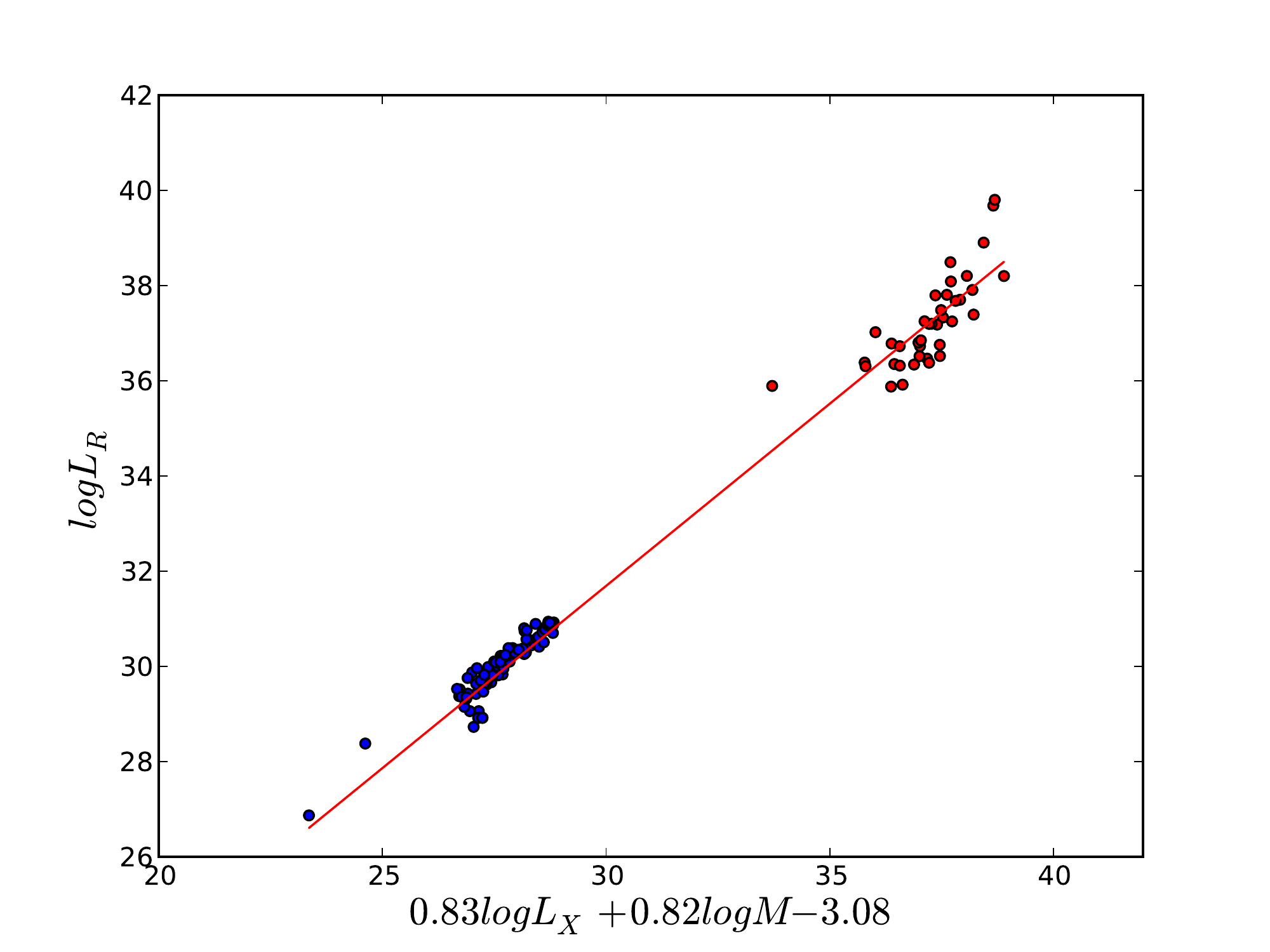}
  \includegraphics[width=87.2mm]{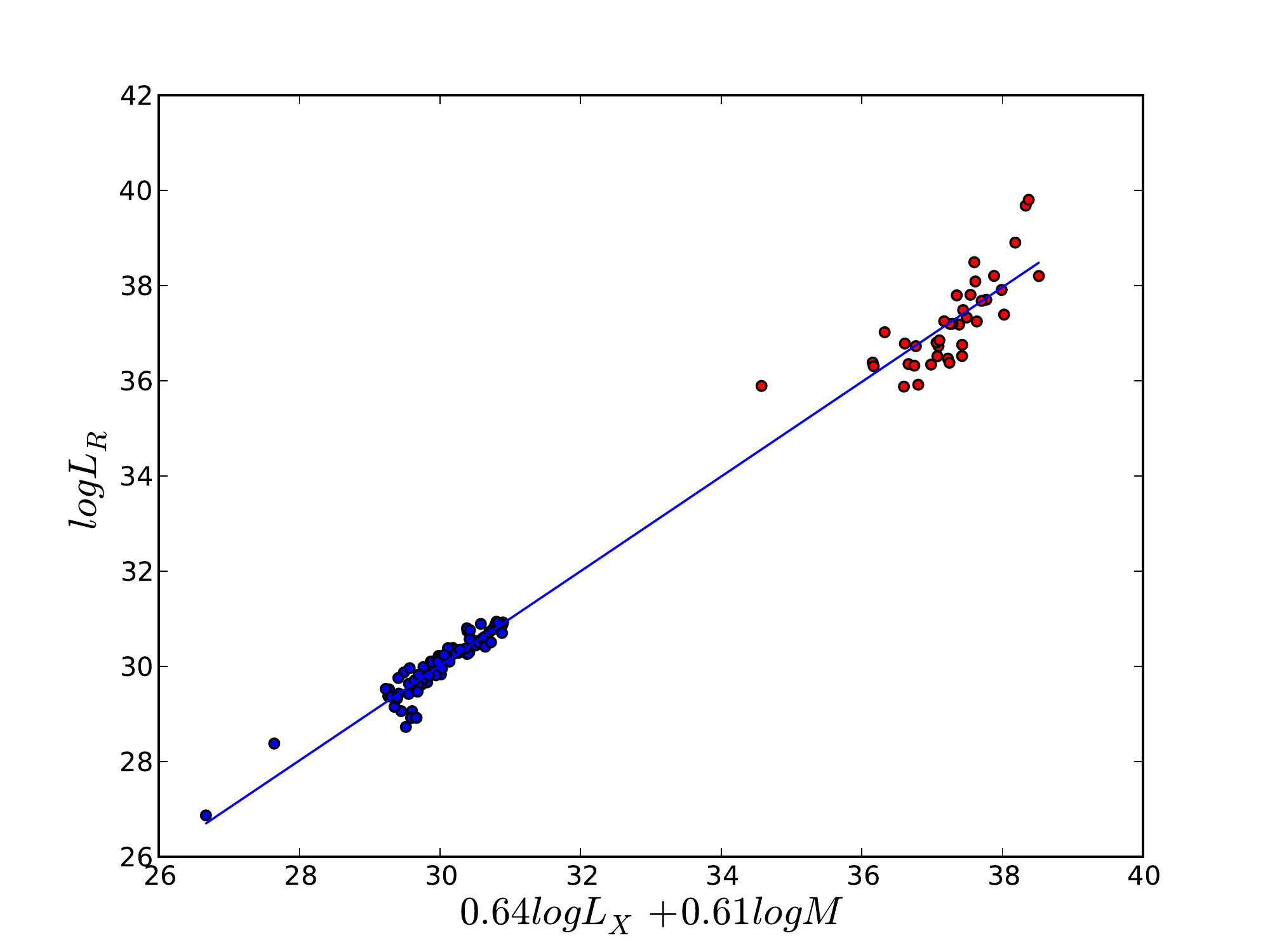}
 \caption{Projection of the Fundamental Plane. In the top plot, SMBH sample is shown in red with the red solid line depicting the best-fit line for the SMBH sample. The XRB sample is put on the graph as blue dots without fitting. In the bottom plot, we show the combined fit for the complete sample including both the SMBH and the XRB sample. The blue line is the obtained best-fit line. Luminosities are given in erg/s while the masses are in the unit of solar mass.}
\end{figure}

\begin{figure}
 \includegraphics[width=87.2mm]{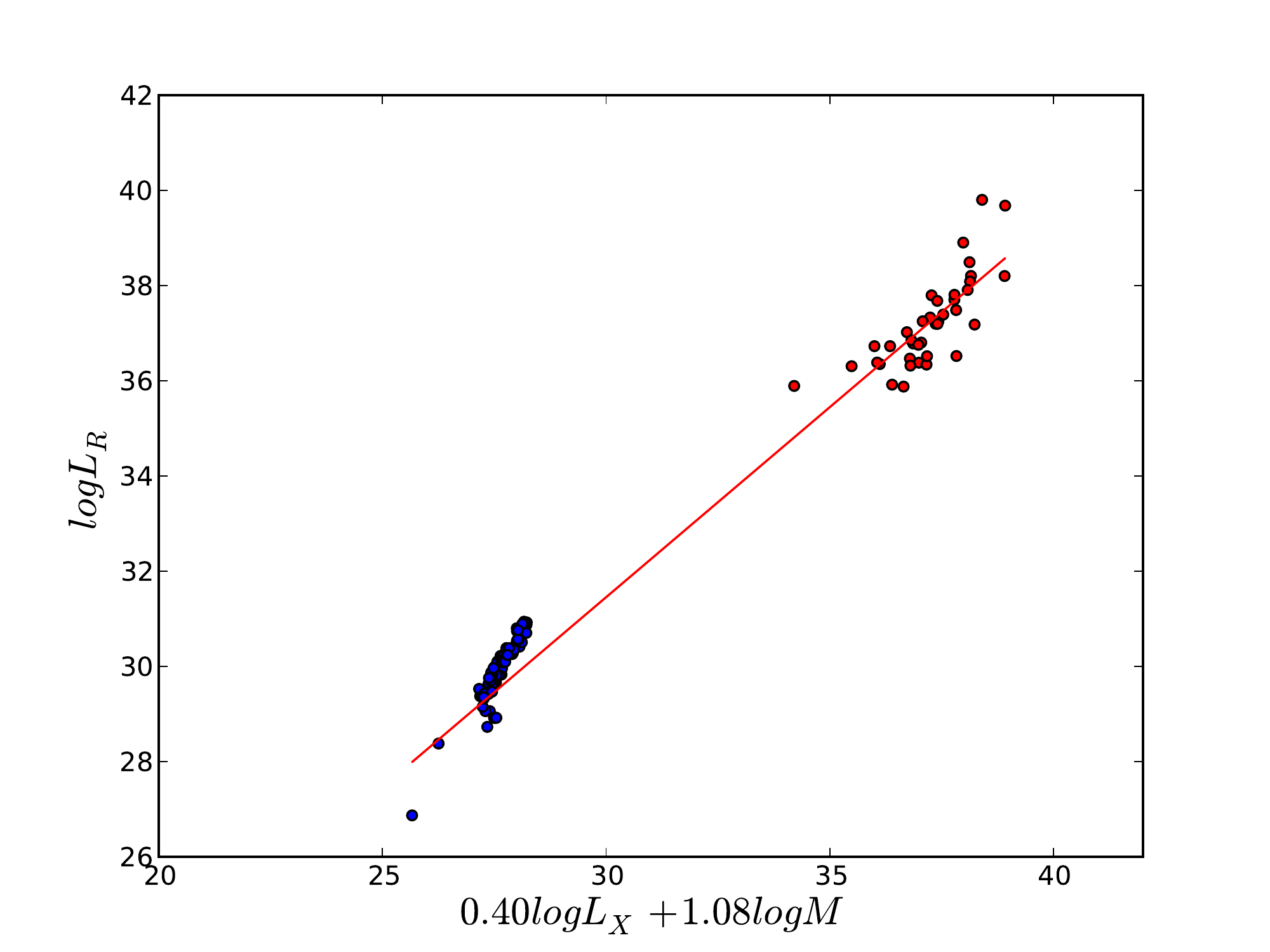}
  \includegraphics[width=87.2mm]{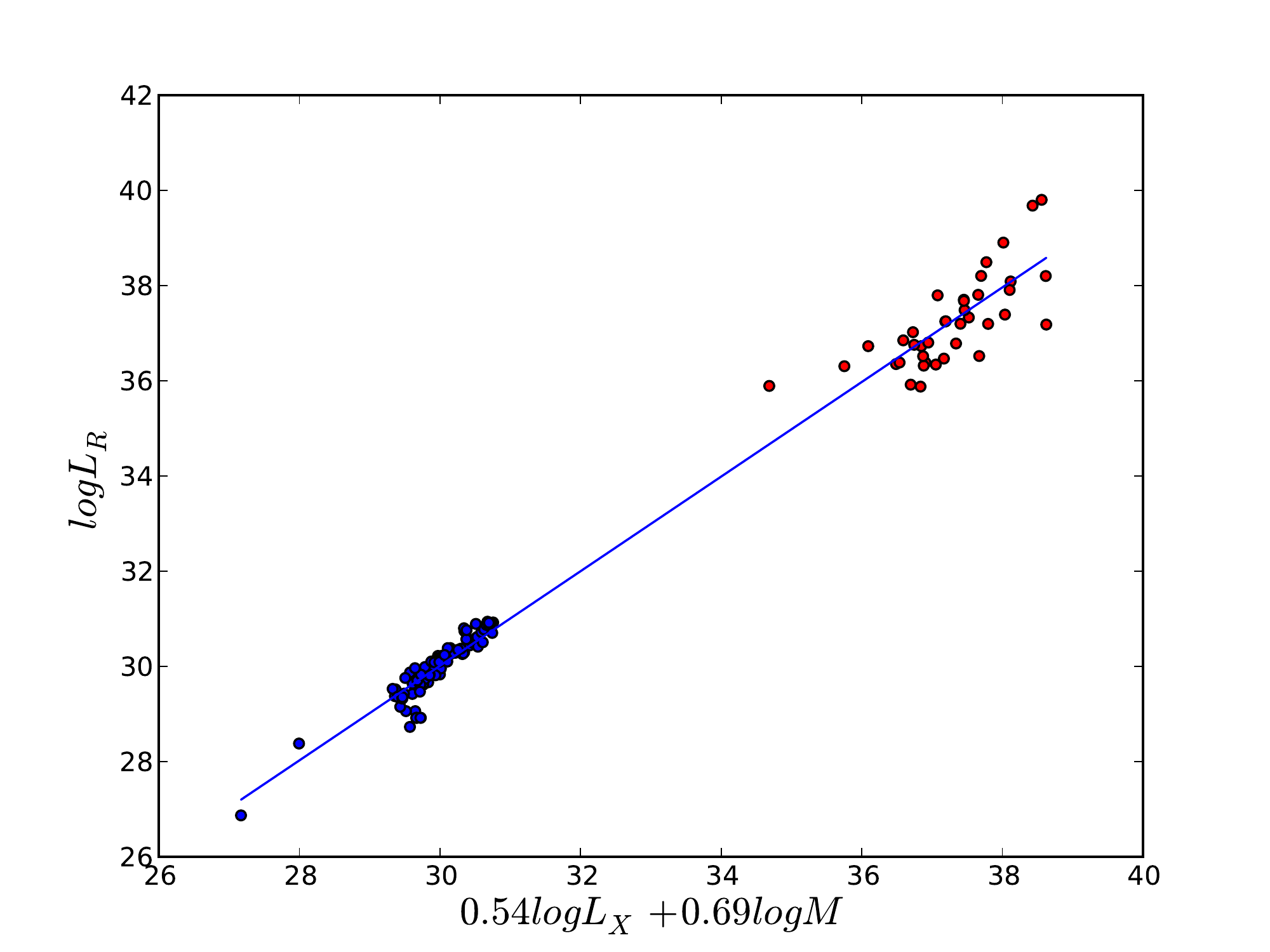}
 \caption{Same as Fig 2, but with X-ray luminosity obtained from the relation proposed by L09.}
 \end{figure}

\subsection{Results}

After statistically verifying the significance of the plane, we use the merit function to estimate the plane equation. We obtain the best fit coefficients for the function 
\begin{center} 
log $L_{R}$ = $\xi_{RO}$ log $L^{c}_{OIII}$ + $\xi_{RM}$ log $M$ + $b_R$ ,
\end{center} 
where $\xi_{RO}$ and $\xi_{RM}$ denote the respective correlation indices and $b_R$ is the constant offset.

We perform the multivariate regression analysis on the complete sample of SMBH using [OIII] line luminosity as an estimate of mass accretion, and obtain the best fit plane described by
\begin{center} 
log $L_{R}$ = (0.83$\pm$0.4) log $L_{OIII}$ + (0.82$\pm$0.3) log $M$.
\end{center}
On the other hand, using the extinction-corrected [OIII] luminosities, we get a plane with different parameters
\begin{center} 
log $L_{R}$ = (0.42$\pm$0.15) log $L^{c}_{OIII}$ + (1.09$\pm$0.22) log $M$
\end{center} 


\section{Reproducing the Fundamental Plane in X-rays}

To reproduce the established Fundamental Plane of Black Hole activity \citep{m03,f04} we have included XRB data to our sample. The fundamental plane of black hole activity is a plane in the space given by black hole mass and the radio/X-ray luminosities, in the form of 
\begin{center} 
log $L_{R}$ =  $\xi_{RX}$ log $L_{X}$ +  $\xi_{RM}$ log $M$ + $b_R$  .
\end{center} 

\subsection{Using observed [OIII] luminosity}

The SMBH [OIII] emission line luminosity is converted to X-ray luminosity with the relation proposed by \cite{h05} (hereafter H05), which states that the hard X-ray (3-20 keV) and [OIII] line luminosities are well-correlated with the mean value for log ($L_{3-20keV}/L_{OIII}$) as 2.15 dex.\\

As the X-ray range used for our study is 2-10 keV, it is required to modify the Heckman relation of  $L_{3-20keV}/L_{OIII}$ to a correlation between luminosities of 2-10 keV and luminosities of [OIII] emission line to ensure consistency.

To find the needed correlation, we use the hard X-ray selected sample of 47 local AGN used in H05. For 23 of these galaxies, 2-10 keV luminosities are provided in H05. 12 more galaxies of the sample have their 2-10keV range luminosities recorded in CAIXA - XMM Newton catalogue \citep{b09}. We compile the luminosities in 2-10 keV range for the complete sample, whenever available. As the photoelectric absorption effects are more significant in the 2-10 keV band compared to the 3-20 keV specially for the Type 2 AGNs which can have high absorbing column densities, we exclude them from the sample. We investigate the correlation of the X-ray (2-10 keV) and [OIII] line luminosities for the Type 1 AGNs of the sample and find the mean value for log ($L_{2-10KeV}/L_{OIII}$) as 1.81 dex. With this conversion factor, we estimate the luminosity of our SMBH sample at 2-10 keV range.\\

\noindent We first fit only the supermassive black hole sample and obtain a relation 
\begin{center} 
log $L_{R}$ = (0.83$\pm$0.3) log $L_{X}$ + (0.82$\pm$0.2) log $M$ - 3.08 .
\end{center}

Extrapolating this relation to lower black hole masses and putting the XRB data in the plot without fitting it, we find that the stellar mass black hole are consistent with the plane found with only the supermassive black holes (see Fig 2). Fitting the XRBs along with the SMBH sample yield results that are in agreement with the relation found with only SMBH, within the errors. This combined fit results in a plane defined by the parameters  $\xi_{RX}$ = 0.64$\pm$0.4 and $\xi_{RM}$= 0.61$\pm$0.2.

This plane is also in agreement with the fundamental plane of black hole activity reported in previous studies. For a sample consisting of XRBs, SgrA, LLAGN, Seyferts and Transitions galaxies, \cite{kfc06} has reported the plane parameters as $\xi_{RX}$ = 0.53$\pm$0.10 and  $\xi_{RM}$ = 0.71$\pm$0.19, while \cite{m03} reports their plane parameters after taking high-state objects into account as  $\xi_{RX}$ = 0.60$\pm$0.11 and  $\xi_{RM}$ = 0.78$\pm$0.10.

\subsection{Using extinction-corrected [OIII] luminosity}


\begin{figure}
 \includegraphics[width=41.9mm]{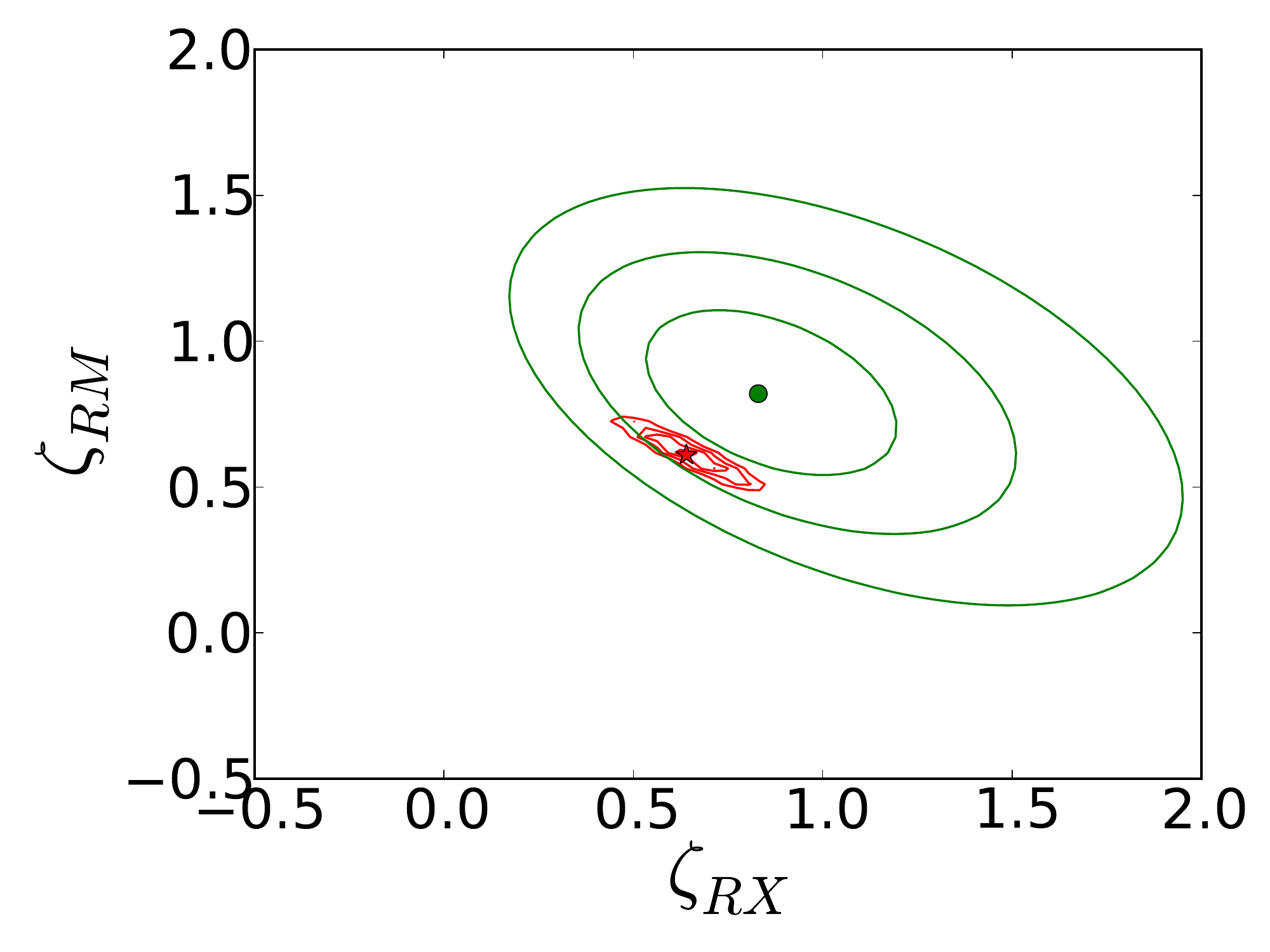}
  \includegraphics[width=41.9mm]{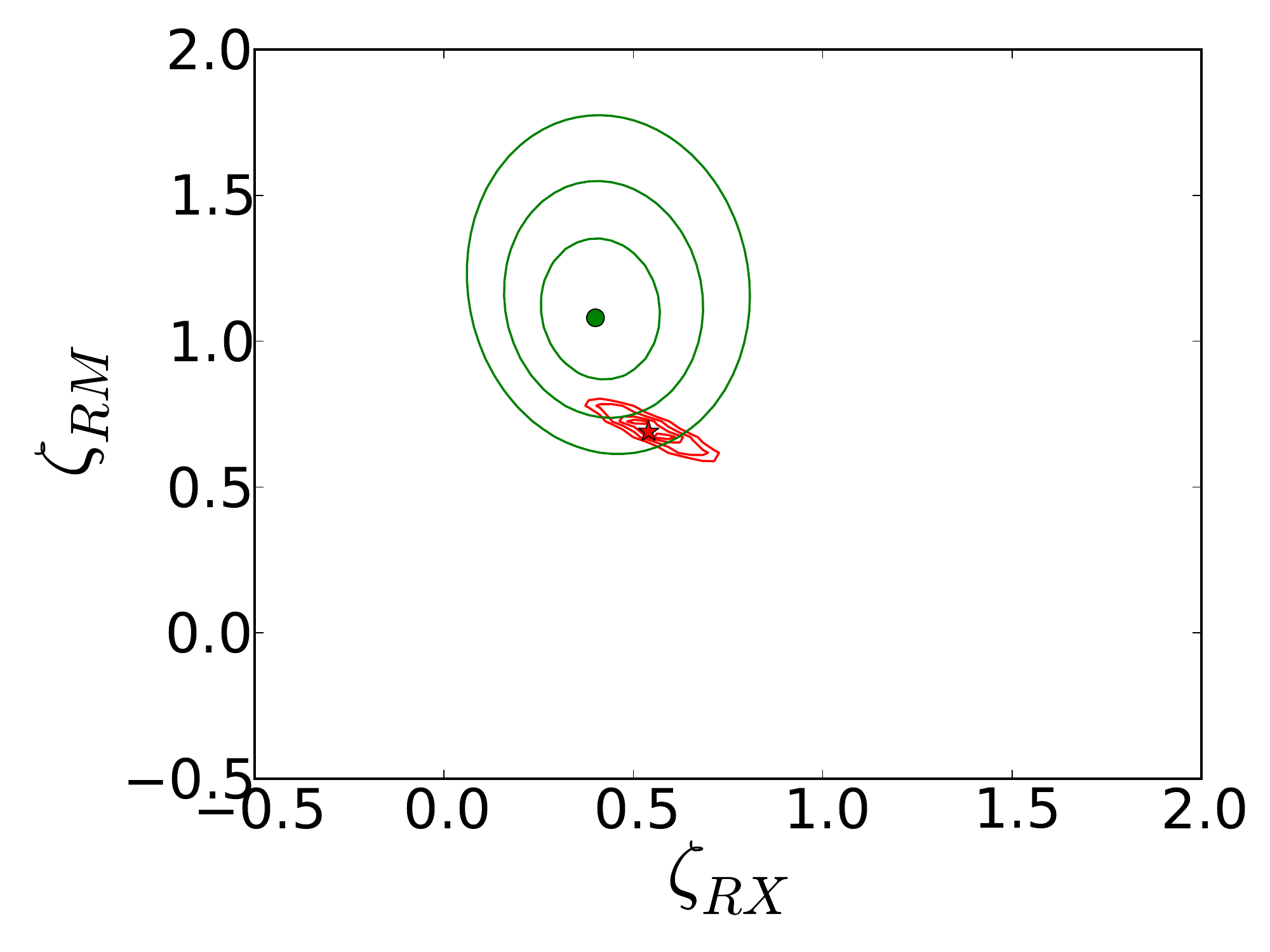}
 \caption{$\chi^2$ confidence ellipses for the observed correlation coefficients $\xi_{RX}$ and $\xi_{RM}$, depicting 1, 2 and 3 sigma confidence levels (Green - only SMBH, Red - both SMBH and XRB). The first plot is for X-ray luminosity obtained from Heckman relation, while the second one is for Lamastra-obtained X-ray luminosity.}
\end{figure}
 
\cite{l09} (hereafter L09) also investigates for a correlation between [OIII] and X-ray luminosities, but unlike H05 they use extinction-corrected [OIII] line luminosities. They report a linear correlation between L$_X$ and L$_{OIII}^c$, which is given by 
\begin{center} 
log $\frac{L_{X}}{10^{42}erg/s}$ = (1.11$\pm$0.10) + (1.02$\pm$0.06) log$\frac{L_{OIII}^c}{10^{42}erg/s}$ .
\end{center}

We see that the correlation coefficients of the plane changes considerably when extinction-corrected [OIII] luminosities are used to obtain the X-ray luminosities. We obtain the best fit coefficients as $\xi_{RX}$  = 0.40$\pm$0.15 and $\xi_{RM}$ = 1.08$\pm$0.22 for the SMBH sample only, and $\xi_{RX}$ = 0.54$\pm$0.1 and  $\xi_{RM}$ = 0.69$\pm$0.1 for the combined fitting of the SMBH and the XRB sample, yielding a plane defined by 
\begin{center} 
log $L_{R}$ = (0.54$\pm$0.1) log $L_{X}$ + (0.69$\pm$0.1) log $M$.
\end{center}

As seen by the bootstrapping error range and chi-square maps (see Fig 4), the fundamental plane obtained from SMBH using extinction-corrected [OIII] luminosities is better-constrained. But strangely in this case the stellar mass black hole data do not agree completely to the fit obtained only with the SMBH sample, although a combined fit of the complete sample does reproduce the fundamental plane. The plots can be seen in Fig 3.\\

In Fig 4, we show our results as $\chi^2$ confidence ellipses traced out by variation of $c$ in the ($a,b $) parameter space. The model used has two degrees of freedom and hence the 1$\sigma$ confidence region is given by $\Delta\chi^2$ = 2.3. Similarly, $\Delta\chi^2$ = 6.18 and $\Delta\chi^2$ = 11.83 represent the 2$\sigma$ and 3$\sigma$ confidence regions, respectively. Here, the contours represent 1, 2 and 3 sigma confidence levels. As shown in the $\chi^2$ confidence maps, while the parameters obtained from fitting only the SMBH gives us a 3$\sigma$ result, the combined fit of the complete sample including both SMBH and XRB gives much better constrained parameters.


\section{Bolometric Fundamental Plane}


\begin{figure}
\vspace{3.02pt}
 \includegraphics[width=41.9mm]{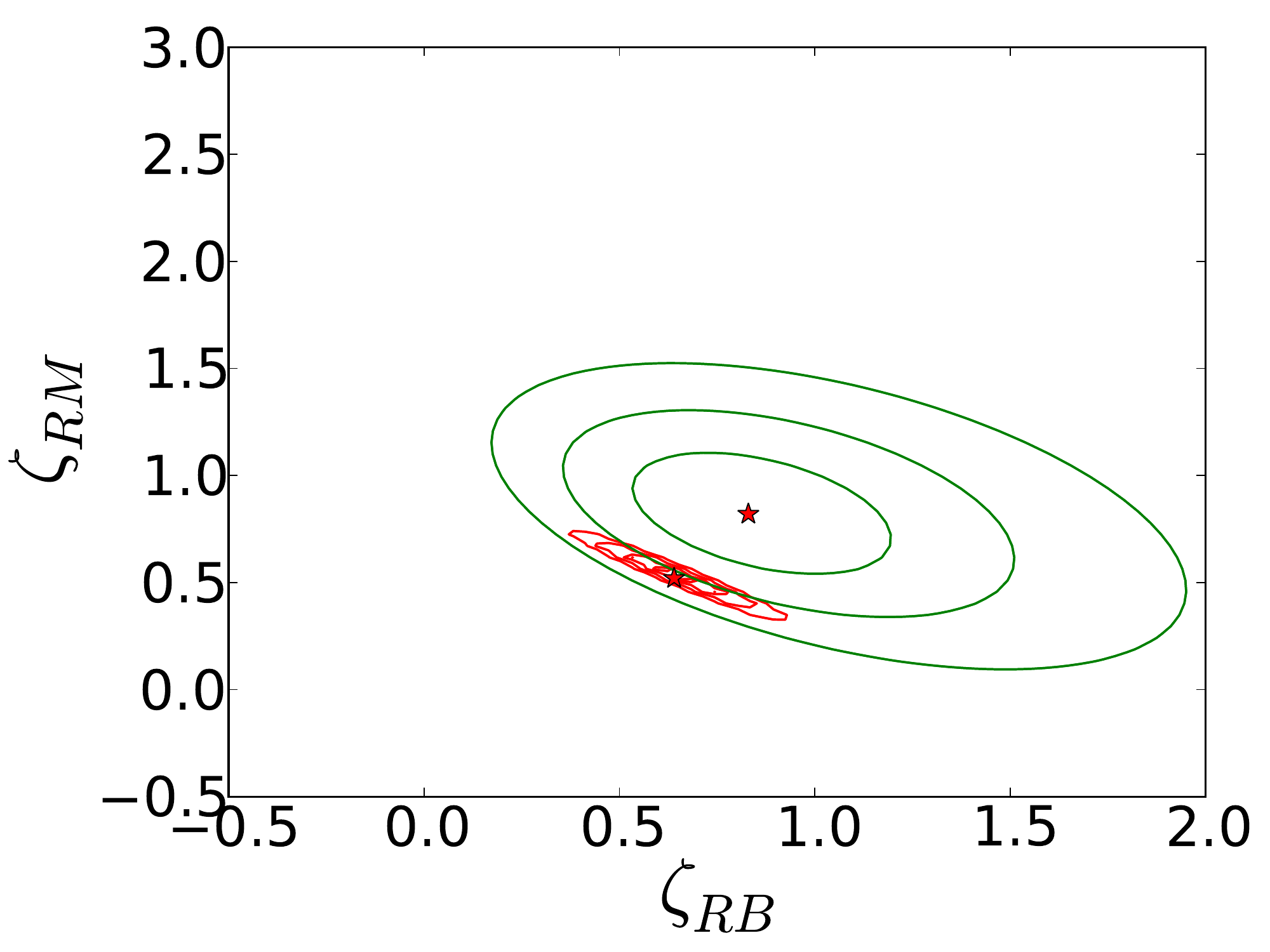}
  \includegraphics[width=41.9mm]{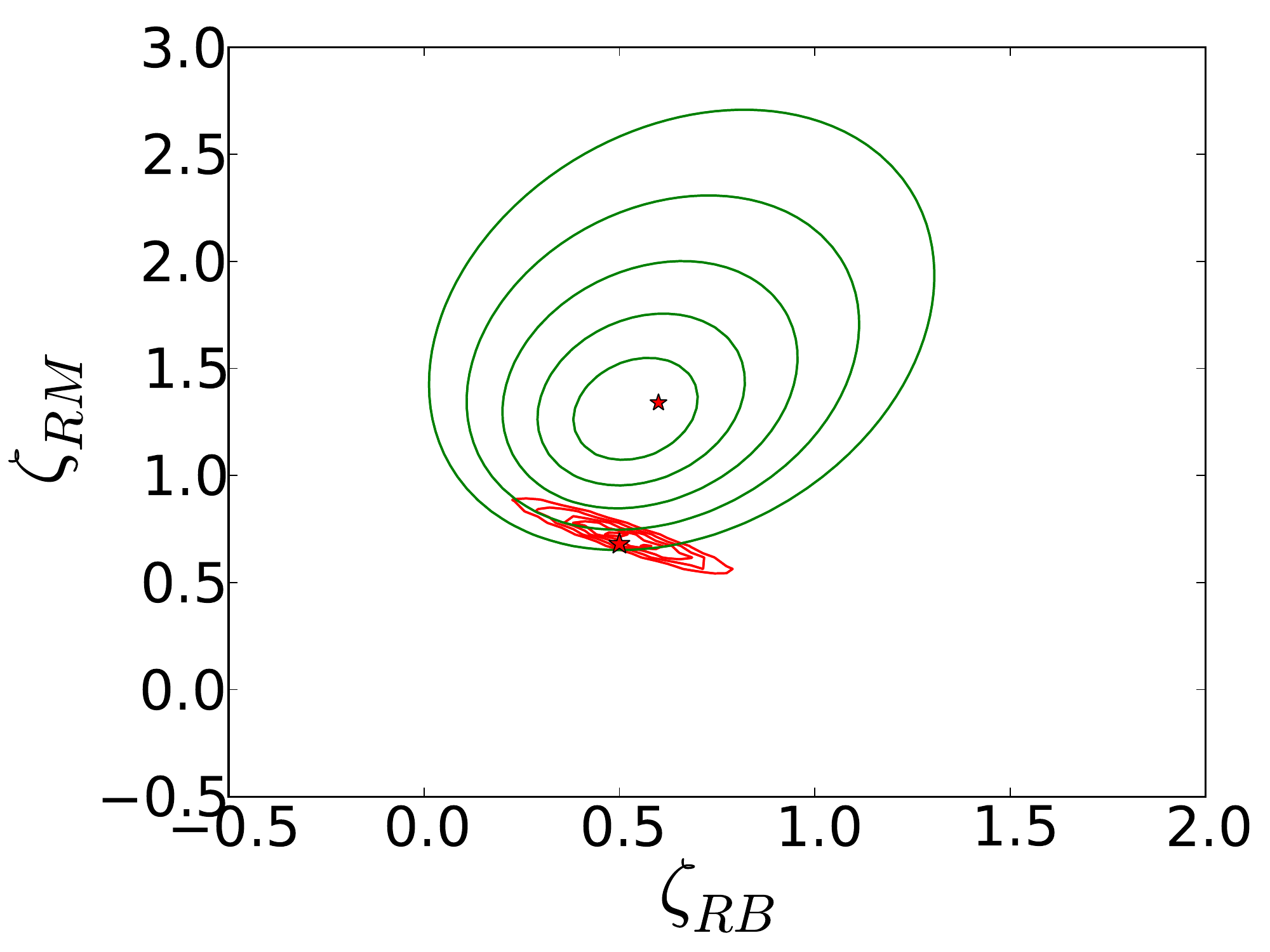}
 \caption{$\chi^2$ confidence ellipses for the observed correlation coefficients $\xi_{RB}$ and $\xi_{RM}$ of the Bolometric fundamental planes (Green - only SMBH, Red - both SMBH and XRB), depicting 1, 2 and 3 sigma confidence levels.The first plot is for X-ray luminosity obtained from Heckman relation, while the second one is for Lamastra-obtained X-ray luminosity.}
\end{figure}


\begin{figure}
 \includegraphics[width=87.2mm]{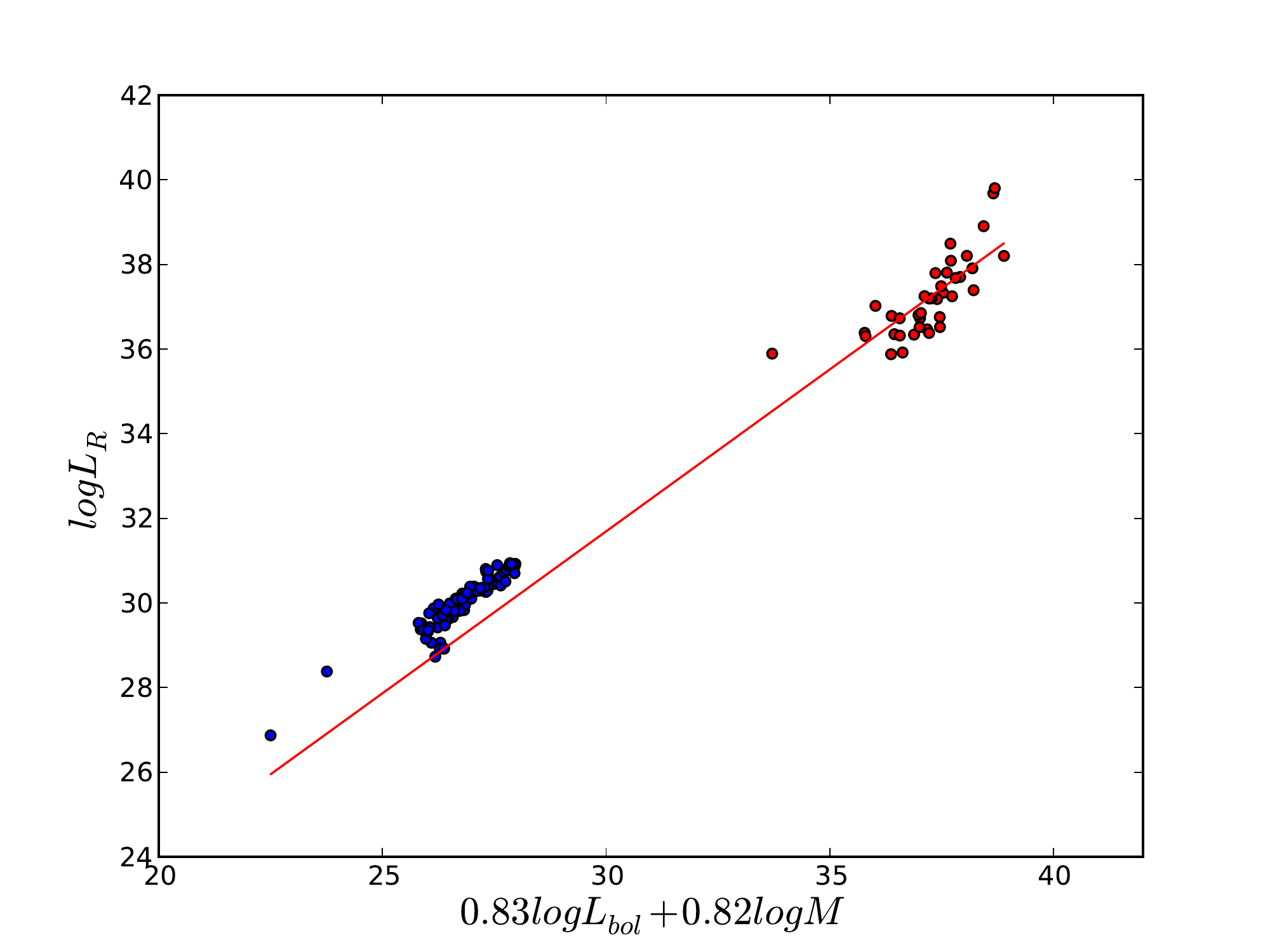}
  \includegraphics[width=87.2mm]{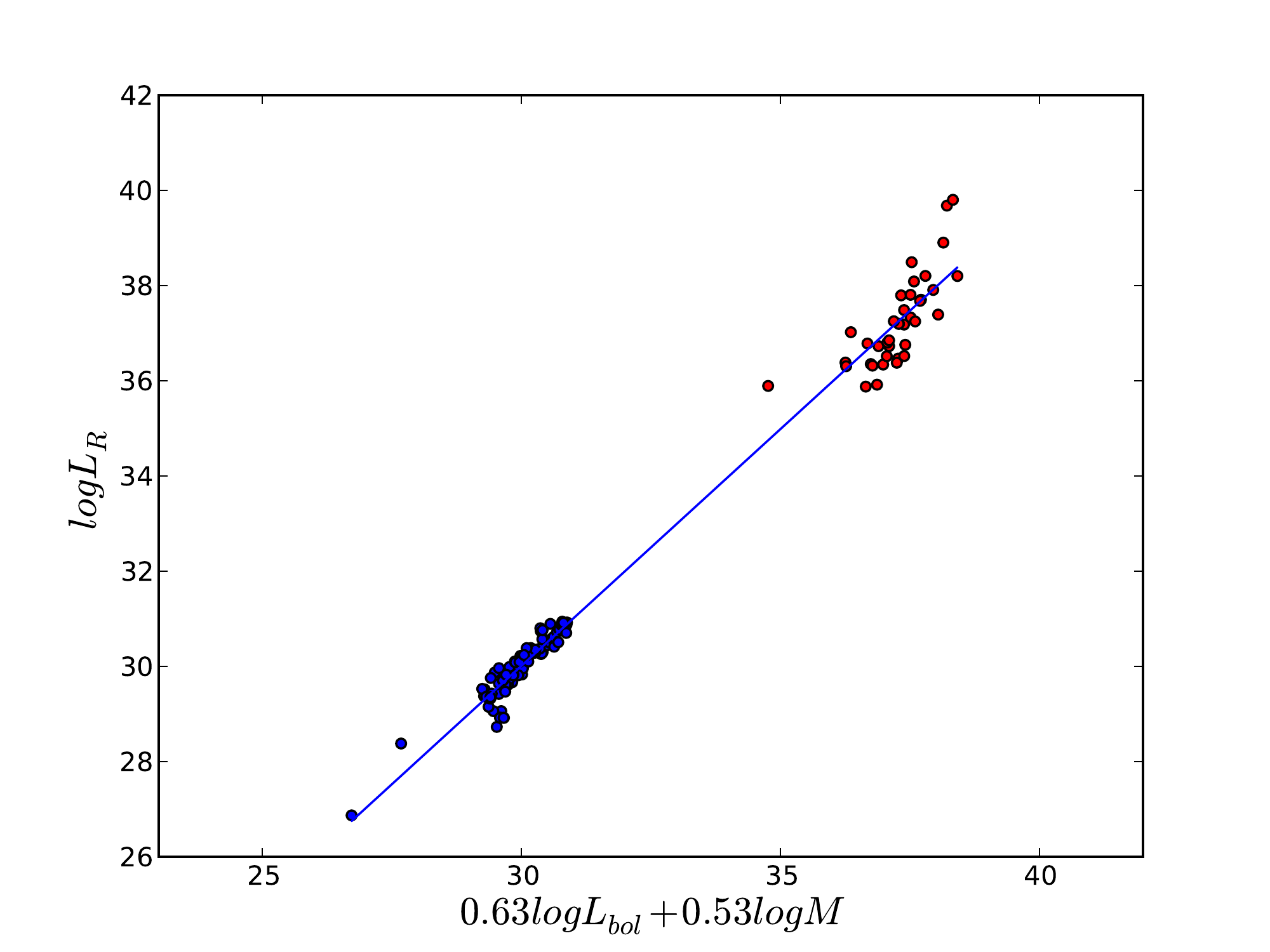}
 \caption{Projection of the Bolometric Fundamental Plane. In the top plot, SMBH sample is shown in red with the red solid line depicting the best-fit line for the SMBH sample. The XRB sample is put on the graph as blue dots without fitting. In the bottom plot, we show the combined fit for the complete sample including both the SMBH and the XRB sample. The blue line is the obtained best-fit line. Luminosities are given in erg/s while the masses are in the unit of solar mass.}
\end{figure}

\begin{figure}
 \includegraphics[width=87.2mm]{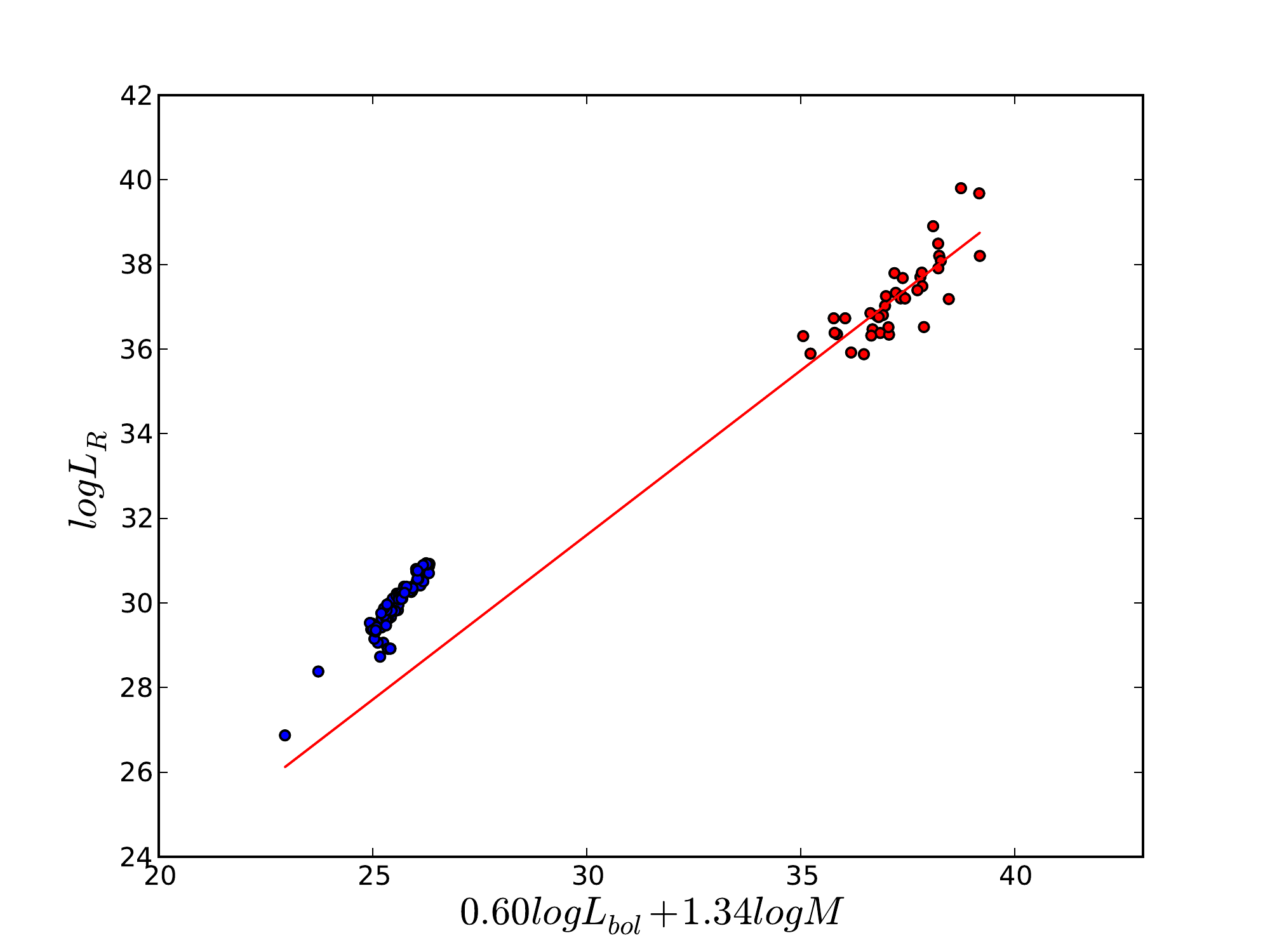}
  \includegraphics[width=87.2mm]{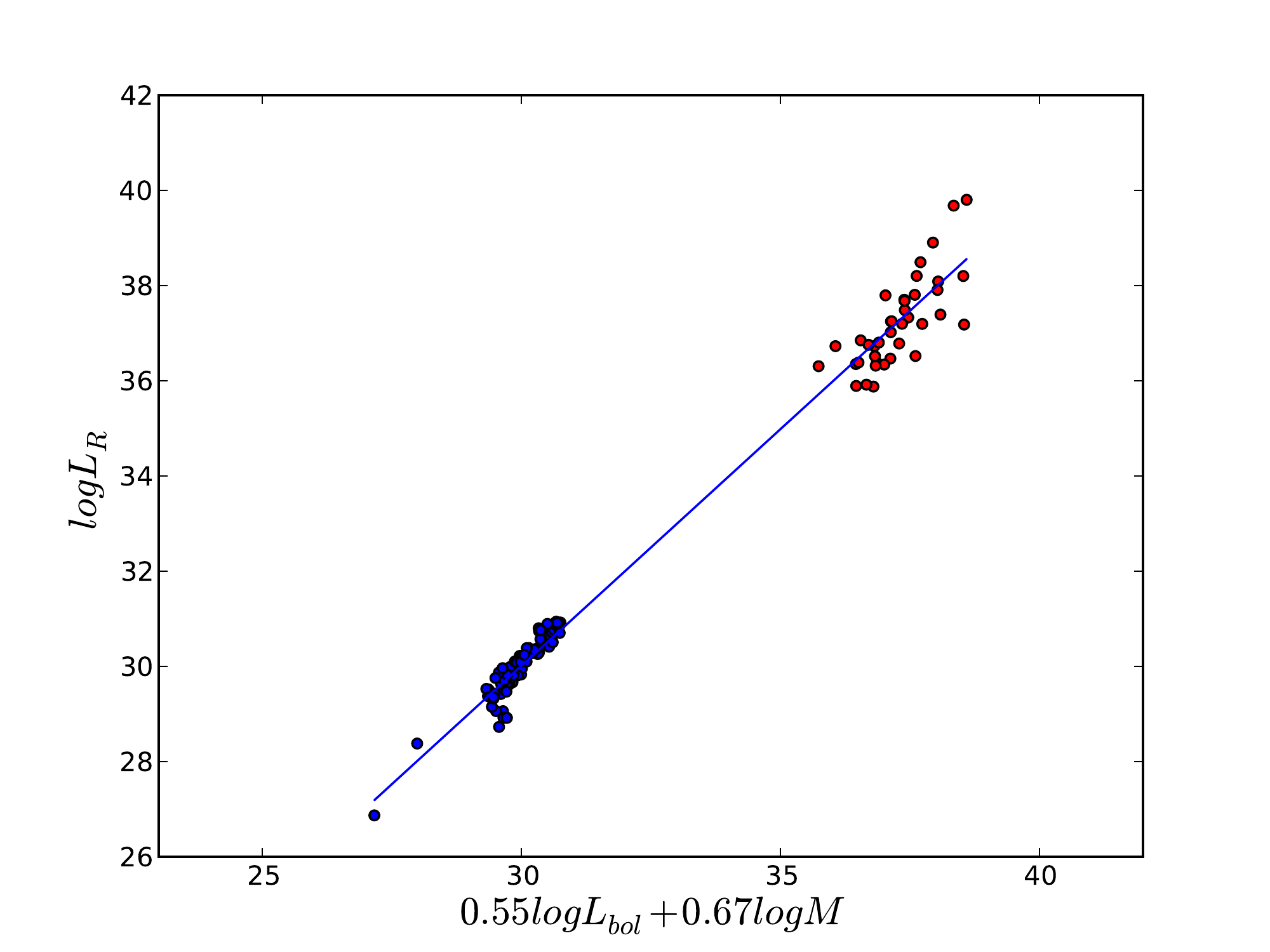}
 \caption{Same as Fig 5, but with Bolometric luminosity obtained from the relation proposed by L09 using extinction-corrected [OIII] luminosity.}
 \end{figure}

To compare the X-ray binaries to our sample of SMBHs, bolometric luminosity can also be used as the estimate for accretion rate. The relation then becomes 
\begin{center} 
log $L_{R}$ = $\xi_{RB}$ log $L_{bol}$ + $\xi_{RM}$ log $M$ + $b_R$ .
\end{center} 

The bolometric luminosity of XRB is taken to be 5 times of the X-ray luminosity. For SMBH, the [OIII] line luminosity can be used as an estimate of the bolometric luminosity of the nuclear source. Different bolometric correction factors have been proposed by different authors.

\cite{h04} has reported the bolometric correction factor as $L_{bol}/L_{OIII}$ $\sim$ 3500, with a variance of 0.38 dex for uncorrected $L_{OIII}$. This relation was obtained using a sample dominated by powerful AGN ($L_{OIII} \sim 10^{6.5}-10^9 L\odot$). They have later extended the sample to lower luminosity AGN and have reported no evidence for systematic luminosity dependence on bolometric correction. But in a more recent study, \cite{l09} has proposed bolometric correction factor to be luminosity dependent. The bolometric correction factor for extinction corrected [OIII] line luminosity is reported to be 87, 142 and 454 for [OIII] emission luminosity ranges of log L$_{OIII}$ (in erg s$^{-1}$) = 38-40,  40-42, 42-44, respectively \citep{l09}.

With these two conversions, we check for the best-fit coefficients for the SMBH sample only and find that while the bolometric luminosity obtained from the Heckman relation yields $\xi_{RB}$ = 0.83$\pm$0.4 and  $\xi_{RM}$ = 0.82$\pm$0.3, the bolometric luminosity calculated with the Lamastra relation results in $\xi_{RB}$ = 0.60$\pm$0.14 and  $\xi_{RM}$ = 1.34$\pm$0.19.  A Kendall Tau correlation test verifies the statistical significance of the plane. For the former plane obtained with Heckman relation, we obtained a Kendall Tau coefficient value of $\tau = 0.56$ (with the probability for null hypothesis as P$_{null} \simeq 5.1 \times 10^{-7})$ while the plane obtained with Lamastra relation gives a Kendall Tau coefficient value of $\tau = 0.66$ (P$_{null} \simeq 2.8 \times 10^{-9})$.

Although a combined fit of the complete sample produces a bolometric fundamental plane relation, but individually the XRBs are incompatible with the bolometric plane stretched out by only the supermassive black holes (see Fig 5 and Fig 6), specially if the conversion of SMBH [OIII] luminosity to bolometric luminosity is obtained from extinction-corrected line luminosity using a luminosity-dependent conversion factor.


\section{Discussion}

\subsection{Recovering the Fundamental Plane of black hole activity}

We recover the fundamental plane of black hole activity with our sample by converting the [OIII] line luminosities to respective X-ray luminosities in the 2-10 keV range. We use two different approaches to convert the [OIII] line luminosities to X-ray (2-10 keV) luminosities - one using the H05 relation and the other using the L09 relation with extinction-corrected [OIII] luminosity. With both these approaches, we perform a combined fit of the supermassive and stellar black hole sample. We find that the plane parameters in both these cases agree to the fundamental plane of black hole activity, within the errors. The parameters for the planes obtained are reported in Table 1.

Using the H05 relation to do the necessary conversion, we also show that the stellar mass black holes completely agree to the best-fit plane of the supermassive sample when extrapolated to lower black hole mass range, even without fitting the stellar mass sample. The combined fit of both supermassive and stellar mass black hole (with H05 conversion) gives a plane
\begin{center}
log $L_{R}$ = (0.64$\pm$0.4) log $L_{X}$ + (0.61$\pm$0.2) log $M$ .
\end{center}
The correlation coefficient found for low/hard state X-ray binaries ($\xi_{RX} \approx$ 0.7) by \cite{gfp03} is consistent within the errors to the value obtained in our combined plane ($\xi_{RX} \approx$ 0.64). This further strengthens our hypothesis that black holes of entire mass range follow these global plane parameters.

\subsection{Discussing the Fundamental Plane in [OIII]}

This study gives a relation between the [OIII] line luminosity of a galaxy hosting a supermassive black hole, it's accretion rate and the mass of the black hole. The plane stretched in the 3D logarithmic space given by the radio and [OIII] luminosity as well as black hole masses follows the relations
\begin{center} 
log $L_R$ = (0.83$\pm$0.4) log $L_{[OIII]}$ + (0.82$\pm$0.3) log $M$ ,
\end{center}
\begin{center} 
log $L_{R}$ = (0.42$\pm$0.2) log $L^c_{[OIII]}$ + (1.09$\pm$0.2) log $M$ .
\end{center}

\begin{table}
 \centering
  \caption{Recovering the Fundamental plane of black hole activity for the complete sample. The values reported are the parameters for the plane relation log $L_{R}$ = $\xi_{Ra}$ log $L_{a}$ + $\xi_{RM}$ log $M$.}
  \begin{tabular}{@{}lrrcc @{}}
     Plane & $\xi_{RX}$ & $\xi_{RM}$  & Kendall $\tau$  & $P_{null}$\\
    \hline
    \hline
     & & & & \\
      Fundamental plane  & &   & &\\
      ($L_{a}$ = $L_{X}$) & &   & &\\
  \hline
 \cite{m03} & 0.60 & 0.78 &  &  \\
 F Plane (H05) & 0.64 & 0.61  & 0.872 &  $< 1\times 10^{-10}$  \\
 F Plane (L09) & 0.54 & 0.69   & 0.877 &  $< 1\times 10^{-10}$ \\
\hline
     & & &  &\\
      Plane in [OIII]  & &   & &\\
      ($L_{a}$ = $L_{OIII}$) &   & &\\
  \hline
[OIII ] Plane & 0.83 & 0.82  & 0.67 & 1.3 $\times 10^{-9}$\\
$[OIII]^c$ Plane & 0.42 & 1.09  & 0.66 & 2.8 $\times 10^{-9}$\\
\hline
     & & &  &\\
      Bolometric Plane & &   & &\\
      ($L_{a}$ = $L_{bol}$) &   & &\\
  \hline
B Plane (H05) & 0.62 & 0.53  & 0.875 &  $< 1\times 10^{-10}$\\
B Plane (L09) & 0.54 & 0.67  & 0.872 &  $< 1\times 10^{-10}$\\
\hline
\end{tabular}
\end{table}

For radiatively inefficient accretion flow, as a first order approximation, analytical scaling of radio luminosity with mass accretion rate is empirically shown to be $L_{R} \propto \dot{M}^{1.4}$ \citep{bk79,fb95,k06}. Using this relation, one can find a rough dependence of [OIII] emission line luminosity on mass accretion for supermassive black holes as
\begin{center} 
$L_{[OIII]} \propto \dot{m}^{1.68} M^{0.69}$ .
\end{center} 
On the other hand, this study shows that the use of extinction-corrected [OIII] luminosity results in a much higher dependence on mass accretion, portraying a relation
\begin{center} 
$L^c_{[OIII]} \propto \dot{m}^{3.33} M^{0.75}$ .
\end{center} 
This relation is in agreement with previous studies showing that at low accretion rate, luminosity of an inefficient accretion structure has steep dependence on $\dot{m}$ \citep{n97,m97}. Recent studies also show that for accretion values lower than the critical value, the ionizing luminosity decreases $\propto \dot{m}^{3.5}$ \citep{s14}.


\begin{figure*}
  \vspace*{4pt}
   \includegraphics[width=58mm]{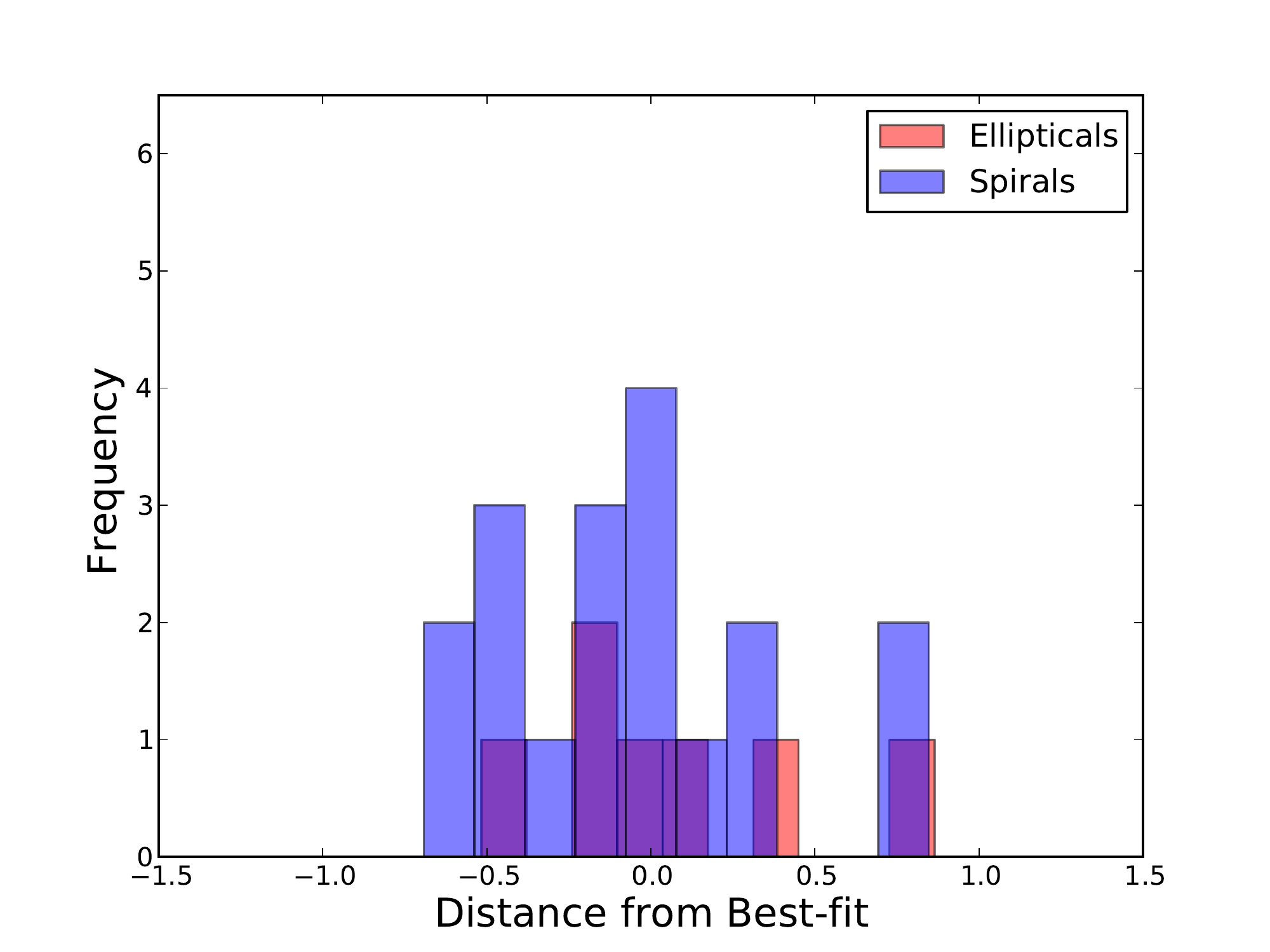}
  \includegraphics[width=58mm]{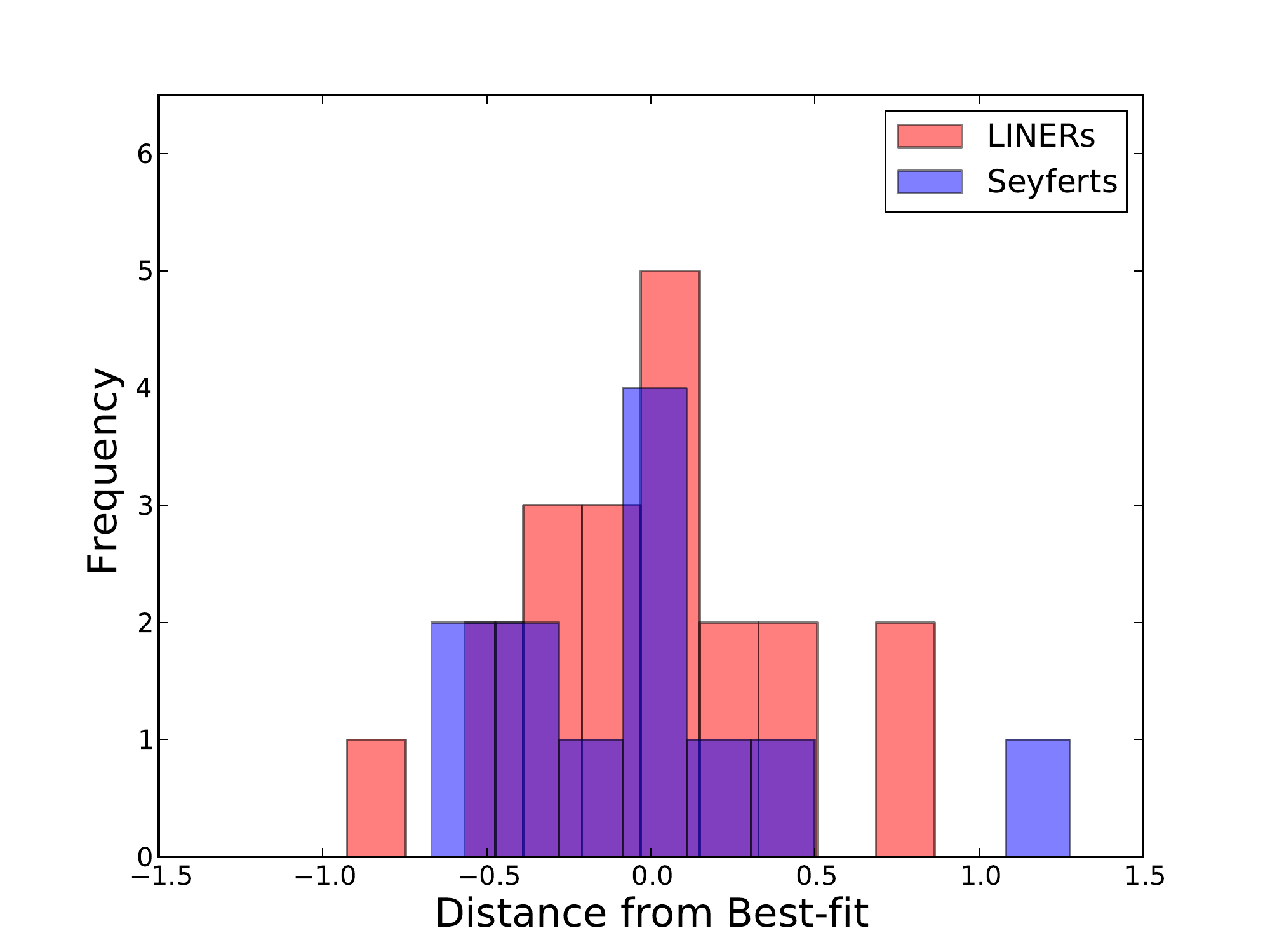}
    \includegraphics[width=58mm]{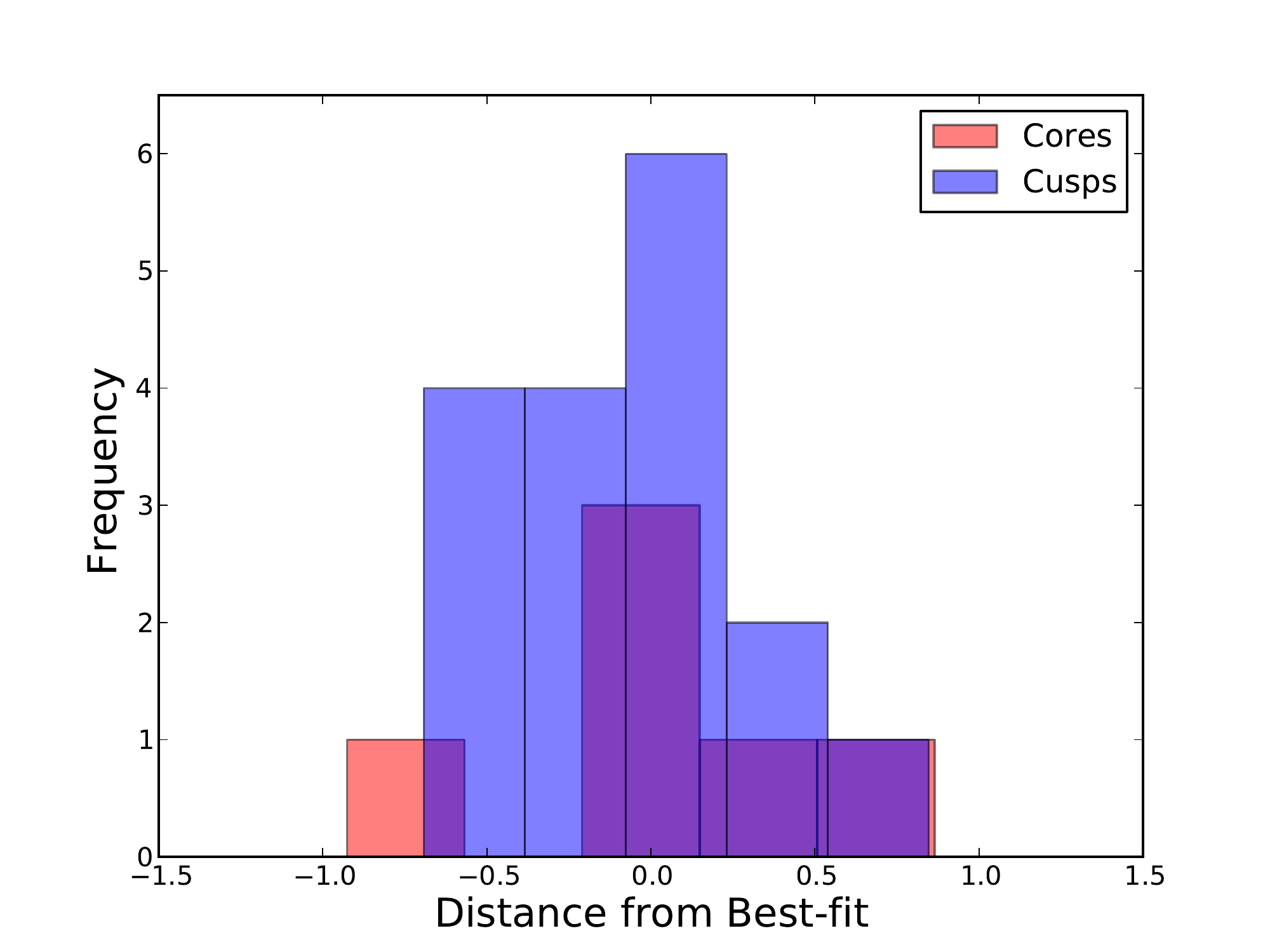}
  \caption{Distances of different classes of AGN from the best-fit plane}
\end{figure*}

\subsection{Introducing a Bolometric fundamental plane}

We also introduce a bolometric fundamental plane obtained by converting the corresponding [OIII] line luminosities to bolometric luminosities. Using a simple linear relation between [OIII] and bolometric luminosities (H05 method), we obtain a plane for the supermassive sample defined by the parameters $\xi_{RB}$ = 0.83$\pm$0.4 and  $\xi_{RM}$ = 0.82$\pm$0.3. Inclusion of the XRBs to the SMBH sample does not change the parameters of the plane considerably. A projection of the plane obtained is shown in Fig 6.

When luminosity dependent bolometric correction factor (L09 method) is used to convert extinction-corrected [OIII] line luminosity to bolometric luminosity, we see that the parameters obtained for the plane are significantly different from the above-mentioned results . The bolometric luminosity obtained using the L09 conversion, depends on the mass accretion rate alone, with almost negligible dependence on black hole mass (see Fig 7). The relation is given by
\begin{center} 
$L_{bol} \propto \dot{m}^{2.62} M^{0.20}$ .
\end{center} 

This surprising negligible mass dependence might be explained as an effect of using luminosity-dependent bolometric correction factors. The bolometric correction factors reported in L09 increases for increasing [OIII] luminosity ranges, which might have effectively changed the mass scaling of bolometric luminosity. The dependency of bolometric luminosity in mass is suspected to be already included in the bolometric correction factor, which explains the negligible dependence of the resulting plane on black hole mass.

\subsection{Radio-loudness in view of the Fundamental Plane}

The Radio-loudness parameter is commonly defined as $R = L_{\nu R}/L_{\nu opt}$, where $L_{\nu R}$ and $L_{\nu opt}$ stand for the monochromatic luminosities at some specified radio and optical frequencies \citep{k89}. Study of this parameter is crucial for addressing basic questions related to formation, acceleration and collimation of jets in different AGN classes. Various studies have shown that different classes of AGN have different Radio-loudness parameters, but whether the radio distribution is continuous or there is a radio bimodality is still a matter of debate \citep[eg.][]{w00}. This dichotomy has been used to infer different physical scenarios - some authors have proposed the Ôspin paradigmÕ, saying that more rapidly spinning black holes result in radio-loud AGNs \citep{b90, wc95, nm12} while some propose the magnetic flux paradigm, where the thin accretion disks give rise to radio-quiet AGN while hot (or thick) accretion disk are more efficient in depositing magnetic flux close to the black hole and hence give rise to radio-loud AGNs \citep{sikora13}.

The diverse sample of LLAGN used for the study follow one single Fundamental Plane equation, with no collective systematic deviation from the plane. As shown in Fig 8, where we have plotted histograms indicating the distances of the different sources from the best-fit plane, we see that no obvious differences can be spotted in the behavior of different AGN classes. Hence, the different types of LLAGN are consistent with having similar physical mechanisms to produce radio luminosity. In view of our plane, radio loudness is a general consequence of difference in Black hole masses - more massive black holes portraying more radio loud galaxies. Hence we suggest that the radio loudness parameter should not be used to infer different physical mechanisms in the sources.

We also perform the Kolmogorov-Smirnov (KS) test to quantify a distance between the empirical distribution functions of both the samples and check if different AGN samples come from the same distribution. We show using KS Statistics that the hypothesis that the underlying distributions of different AGN samples are identical can not be rejected.

\subsubsection{ Ellipticals and Spirals}

We construct the fundamental plane separately for the ellipticals (sample - 7 ellipticals) and the spirals (sample - 18 spirals) and find the slopes to agree with each other within the uncertainties. A Kolmogorov-Smirnov (KS) test on the samples shows that the underlying distribution of the two samples are in agreement with being similar (KS statistics = 0.21, p-value = 0.95).

\subsubsection{ LINERs and Seyferts}

By fitting the plane separately, we see that both the LINERs (sample - 20 LINERs) and Seyferts (sample - 12 seyferts) hold very similar Fundamental Plane relations. The histograms showing the deviation of these sources from the fundamental plane can be seen in Fig 8. We also calculate the distances of the individual sources from the best-fit plane and compare the distribution functions of the LINERs and Seyferts. We see that both the types of AGN follow similar distribution functions. We get the KS statistics value as 0.25 and the p-value as 0.67. Hence, the hypothesis that the underlying distributions of the LINER and seyfert samples are identical can not be rejected.

\subsubsection{ Core, S\'{e}rsic and Double-S\'{e}rsic Galaxies}

The brightness profile of the central regions of galaxies are described by the Nuker law \citep{l95}. Galaxies can be divided into two distinct classes according to their central brightness profile - the steep power law galaxies and the core galaxies with flattened slope. An alternate and more powerful description of this core/power-law dichotomy is based on inner logarithmic slope and is known as the S\'{e}rsic/core-S\'{e}rsic classification. The nuclei of core galaxies are considered to be radio-loud while those of cusp galaxies are radio-quiet \citep{cv06,ruk11}.

In our sample, we find that the core galaxies stretch out in a plane with a slope of 0.78, while both the S\'{e}rsic and double-S\'{e}rsic subsamples have similar slope of 0.95. It is interesting to compare the slight deviations in slope of the core and S\'{e}rsic subsamples as well as the elliptical and spiral subsamples. While both the ellipticals and the S\'{e}rsic samples has similar slopes of 0.7, both the spirals and the core galaxies has a higher slope of 0.9. A nonparametric KS test performed on the core and the cusp (S\'{e}rsic as well as double-S\'{e}rsic) galaxies showed that the underlying distribution of both the samples can be same (KS statistics = 0.27, p-value = 0.83). 


\section[]{Conclusion}

We use a complete sample of 39 supermassive black holes selected from the Palomar Spectroscopic Survey and a selected sample of the best-studied stellar mass X-ray binaries, to study the general properties of accretion. We present results of multivariate regression analysis performed on the same and show that

\begin{itemize}
\item A fundamental plane of black hole activity is seen in the logarithmic 3-dimensional space of black hole mass, radio luminosity and [OIII] emission line luminosity.
\item The established fundamental plane of black hole activity in X-rays can be reproduced with this sample. It is possible to obtain the plane relation individually by the supermassive sample alone. On extrapolation to the lower mass black hole range, we see that the X-ray binaries agree to the plane stretched by the supermassive sample.
\item We also introduce a bolometric fundamental plane for the supermassive sample, which does not completely agree to the complete sample after inclusion of the X-ray binaries. This plane shows negligible dependence on black hole mass, which might be due to the use of luminosity-dependent bolometric correction factor that effectively corrects for mass-scaling of bolometric luminosity.
\item Finally, we look at the implication of the plane in radio-loudness of different AGN types. We see that after accounting for the non-linearity in the radio-[OIII] luminosity correlation and by including a mass scaling factor, we see no clear radio-dichotomy in the different types of AGN in our sample. 
\end{itemize}


\bsp


\appendix

\section{Main properties of the sample}

\begin{table}
\centering
  \caption{SMBH Sample : Main properties}
\begin{tabular}{lrrr}
\hline
 Name    &        Mass (in $M_{\odot}$) &       $L_R$ (in erg/s) &   $L_{OIII}$ (in erg/s)  \\
\hline
 NGC5033 & 2.89523e+07 & 2.92753e+36 &  7.46678e+39 \\
 NGC4138 & 1.48166e+07 & 2.24663e+36 &  1.04037e+39 \\
 NGC3998 & 1.8723e+09  & 1.59028e+38 &  1.62112e+40 \\
 NGC2655 & 6.18073e+07 & 2.1361e+37  &  1.24977e+40 \\
 NGC5353 & 1.43127e+09 & 1.59777e+38 &  5.13102e+38 \\
 NGC5354 & 2.65312e+08 & 6.24036e+37 &  3.26808e+38 \\
 NGC4278 & 6.72349e+08 & 5.04692e+37 &  4.76834e+38 \\
 NGC3226 & 3.51531e+08 & 1.76814e+37 &  3.60821e+38 \\
 NGC4419 & 2.32308e+07 & 6.07591e+36 &  1.99971e+40 \\
 NGC4486 & 3.43764e+09 & 4.78596e+39 &  3.51229e+39 \\
 NGC5846 & 7.61949e+08 & 3.05999e+37 &  4.26385e+38 \\
 NGC2273 & 4.30474e+07 & 2.45978e+37 &  1.63045e+41 \\
 NGC6500 & 2.77027e+08 & 8.01106e+38 &  1.45928e+40 \\
 NGC3190 & 1.42429e+08 & 2.40036e+36 &  3.49149e+38 \\
 NGC5363 & 3.75418e+08 & 1.22118e+38 &  1.53139e+40 \\
 NGC3147 & 3.16161e+08 & 8.10255e+37 &  1.79206e+40 \\
 NGC3628 & 1.50039e+06 & 7.79999e+35 &  1.06867e+37 \\
 NGC3627 & 4.43963e+07 & 7.55398e+35 &  1.11365e+39 \\
 NGC7743 & 3.14113e+06 & 5.34024e+36 &  3.14872e+40 \\
 NGC4169 & 1.2373e+08  & 1.51898e+37 &  4.93036e+41 \\
 NGC3169 & 1.44378e+08 & 1.57809e+37 &  2.65927e+39 \\
 NGC4203 & 1.22352e+08 & 5.34512e+36 &  1.46688e+37 \\
 NGC4293 & 1.01086e+07 & 2.41945e+36 &  2.05736e+39 \\
 NGC4374 & 1.00496e+09 & 3.1004e+38  &  1.08288e+39 \\
 NGC4579 & 1.44378e+08 & 4.77634e+37 &  3.26039e+39 \\
 NGC7479 & 5.235e+07   & 1.56935e+37 &  4.76577e+40 \\
 NGC5377 & 8.40809e+07 & 1.78146e+37 &  2.21031e+39 \\
 NGC4548 & 1.06744e+07 & 2.0253e+36  &  7.56412e+37 \\
 NGC4258 & 2.40533e+07 & 8.29525e+35 &  1.34388e+39 \\
 NGC3607 & 4.19661e+08 & 3.31531e+36 &  2.11336e+39 \\
 NGC5866 & 8.25699e+07 & 1.04987e+37 &  3.29625e+38 \\
 NGC4216 & 1.81405e+08 & 2.19408e+36 &  4.72302e+38 \\
 NGC4589 & 3.67869e+08 & 6.40442e+37 &  2.34894e+39 \\
 NGC4565 & 7.16962e+07 & 2.08178e+36 &  7.39625e+38 \\
 NGC1167 & 2.95376e+08 & 6.34151e+39 &  1.27083e+41 \\
 NGC3945 & 1.5611e+08  & 6.35733e+36 &  3.65437e+38 \\
 NGC2787 & 1.86646e+08 & 7.07416e+36 &  6.76e+37    \\
 NGC4143 & 2.2071e+08  & 5.70298e+36 &  1.0395e+38  \\
 NGC4636 & 3.26608e+08 & 3.28353e+36 &  1.07745e+38 \\
\hline
\end{tabular}
\end{table}

\begin{table}
\centering
  \caption{XRB Sample : Main properties}
\begin{tabular}{lrrr}
\hline
 Name    &        Mass (in $M_{\odot}$) &       $L_R$ (in erg/s) &   $L_{X}$ (in erg/s)  \\
\hline
 GX 339-4      &   0.85 & 30.2947 & 36.8804 \\
 GX 339-4      &   0.85 & 30.2964 & 36.8865 \\
 GX 339-4      &   0.85 & 30.2214 & 36.4579 \\
 GX 339-4      &   0.85 & 30.2253 & 36.3264 \\
 GX 339-4      &   0.85 & 29.6867 & 35.7584 \\
 GX 339-4      &   0.85 & 29.8308 & 36.2568 \\
 GX 339-4      &   0.85 & 30.2596 & 36.8415 \\
 GX 339-4      &   0.85 & 30.2748 & 36.8639 \\
 GX 339-4      &   0.85 & 30.07   & 36.1511 \\
 GX 339-4      &   0.85 & 30.0423 & 36.1014 \\
 GX 339-4      &   0.85 & 30.385  & 36.5264 \\
 GX 339-4      &   0.85 & 29.6798 & 35.7412 \\
 GX 339-4      &   0.85 & 30.4414 & 37.0319 \\
 GX 339-4      &   0.85 & 30.3732 & 36.8019 \\
 GX 339-4      &   0.85 & 29.9292 & 36.1871 \\
 GX 339-4      &   0.85 & 29.5174 & 35.104  \\
 GX 339-4      &   0.85 & 30.2887 & 36.6635 \\
 GX 339-4      &   0.85 & 29.6657 & 35.9521 \\
 GX 339-4      &   0.85 & 29.6163 & 35.8259 \\
 GX 339-4      &   0.85 & 29.4217 & 35.541  \\
 GX 339-4      &   0.85 & 29.6243 & 35.8174 \\
 GX 339-4      &   0.85 & 29.06   & 35.6163 \\
 GX 339-4      &   0.85 & 29.3822 & 35.2641 \\
 GX 339-4      &   0.85 & 29.3753 & 35.088  \\
 GX 339-4      &   0.85 & 29.6282 & 35.5829 \\
 GX 339-4      &   0.85 & 29.6359 & 35.5467 \\
 GX 339-4      &   0.85 & 29.06   & 35.3753 \\
 GX 339-4      &   0.85 & 29.5274 & 35.03   \\
 GX 339-4      &   0.85 & 30.0483 & 36.1957 \\
 GX 339-4      &   0.85 & 30.0028 & 36.0686 \\
 GX 339-4      &   0.85 & 30.2204 & 36.2061 \\
 GX 339-4      &   0.85 & 29.7352 & 35.8662 \\
 GX 339-4      &   0.85 & 29.8547 & 35.9808 \\
 GX 339-4      &   0.85 & 29.8209 & 35.9521 \\
 GX 339-4      &   0.85 & 29.986  & 36.0798 \\
 GX 339-4      &   0.85 & 30.9233 & 37.6382 \\
 GX 339-4      &   0.85 & 30.8729 & 37.623  \\
 GX 339-4      &   0.85 & 29.1511 & 35.2263 \\
 GX 339-4      &   0.85 & 29.6163 & 35.7472 \\
 GX 339-4      &   0.85 & 30.5381 & 37.099  \\
 GX 339-4      &   0.85 & 30.4988 & 37.1011 \\
 GX 339-4      &   0.85 & 30.5906 & 37.1851 \\
 GX 339-4      &   0.85 & 30.6183 & 37.2245 \\
 GX 339-4      &   0.85 & 30.7215 & 37.3335 \\
 GX 339-4      &   0.85 & 30.7749 & 37.4036 \\
 GX 339-4      &   0.85 & 30.8725 & 37.4639 \\
 GX 339-4      &   0.85 & 30.9385 & 37.4921 \\
 GX 339-4      &   0.85 & 30.9243 & 37.5183 \\
 GX 339-4      &   0.85 & 30.858  & 37.5309 \\
 GX 339-4      &   0.85 & 30.2596 & 36.5293 \\
 GX 339-4      &   0.85 & 30.4135 & 37.243  \\
 GX 339-4      &   0.85 & 30.7031 & 37.6152 \\
 GX 339-4      &   0.85 & 30.5066 & 37.37   \\
 GX 339-4      &   0.85 & 29.361  & 35.1534 \\
 GX 339-4      &   0.85 & 29.428  & 35.3232 \\
 GX 339-4      &   0.85 & 29.4694 & 35.7376 \\
 GX 339-4      &   0.85 & 29.3232 & 35.2818 \\
 GX 339-4      &   0.85 & 29.3537 & 35.2818 \\
 GX 339-4      &   0.85 & 29.9558 & 36.2818 \\
 GX 339-4      &   0.85 & 30.1014 & 36.4486 \\
 GX 339-4      &   0.85 & 29.8133 & 36.0078 \\
 GX 339-4      &   0.85 & 29.8133 & 36.1416 \\
 GX 339-4      &   0.85 & 29.8308 & 35.9911 \\
 \hline
\end{tabular}
\end{table}

 \begin{table}
\centering
  \caption{... continued}
\begin{tabular}{lrrr}
\hline
 Name    &        Mass (in $M_{\odot}$) &       $L_R$ (in erg/s) &   $L_{X}$ (in erg/s)  \\
\hline
 GX 339-4      &   0.85 & 29.8108 & 35.9843 \\
 GX 339-4      &   0.85 & 30.3661 & 36.5224 \\
 GX 339-4      &   0.85 & 28.729  & 35.4783 \\
 GX 339-4      &   0.85 & 29.9877 & 35.8689 \\
 GX 339-4      &   0.85 & 30.1027 & 36.0331 \\
 GX 339-4      &   0.85 & 30.1683 & 36.1914 \\
 GX 339-4      &   0.85 & 30.1925 & 36.2972 \\
 GX 339-4      &   0.85 & 30.2312 & 36.3931 \\
 GX 339-4      &   0.85 & 30.9112 & 37.5736 \\
 GX 339-4      &   0.85 & 30.9078 & 37.5395 \\
 GX 339-4      &   0.85 & 30.9173 & 37.5294 \\
 GX 339-4      &   0.85 & 29.7034 & 35.6692 \\
 GX 339-4      &   0.85 & 29.8259 & 35.7647 \\
 XTE J1118+480 &   1    & 28.92   & 35.57   \\
 XTE J1118+480 &   1    & 28.92   & 35.46   \\
 XTE J1118+480 &   1    & 28.92   & 35.47   \\
 XTE J1118+480 &   1    & 28.92   & 35.45   \\
 XTE J1118+480 &   1    & 28.92   & 35.56   \\
 V404 Cyg      &   1    & 30.5637 & 36.7929 \\
 V404 Cyg      &   1    & 28.3794 & 32.4075 \\
 V404 Cyg      &   1    & 30.5782 & 36.7841 \\
 V404 Cyg      &   1    & 30.2842 & 36.4424 \\
 V404 Cyg      &   1    & 30.7422 & 36.7018 \\
 V404 Cyg      &   1    & 30.0895 & 36.0074 \\
 V404 Cyg      &   1    & 30.8014 & 36.6908 \\
 V404 Cyg      &   1    & 29.872  & 35.2893 \\
 V404 Cyg      &   1    & 30.173  & 36.0732 \\
 V404 Cyg      &   1    & 30.1151 & 36.0067 \\
 V404 Cyg      &   1    & 30.3491 & 36.5397 \\
 V404 Cyg      &   1    & 30.3839 & 36.2693 \\
 V404 Cyg      &   1    & 30.0851 & 35.922  \\
 V404 Cyg      &   1    & 30.0939 & 36.0506 \\
 V404 Cyg      &   1    & 30.571  & 36.7517 \\
 V404 Cyg      &   1    & 30.8932 & 36.9999 \\
 V404 Cyg      &   1    & 29.7566 & 35.1663 \\
 V404 Cyg      &   1    & 29.9631 & 35.4186 \\
 V404 Cyg      &   1    & 30.24   & 36.1864 \\
 V404 Cyg      &   1    & 30.7613 & 36.7581 \\
 AO6200        &   1.04 & 26.87   & 30.85   \\
\hline
\end{tabular}
\end{table}

\label{lastpage}

\end{document}